\documentclass{aastex}

\usepackage{emulateapj5,apjfonts}

\begin{document}


\title{A Black Hole in the Superluminal source SAX J1819.3-2525
(V4641 Sgr)$^1$}
%

\author{Jerome A. Orosz}
\affil{Astronomical Institute, Utrecht University, Postbus 80000,
3508 TA Utrecht, The Netherlands}
\email{J.A.Orosz@astro.uu.nl}

\author{Erik Kuulkers\altaffilmark{2}}
\affil{Space Research Organization Netherlands,
Sorbonnelaan 2,
3584 CA Utrecht, The Netherlands}
\email{e.kuulkers@sron.nl}

\author{Michiel van der Klis}
\affil{Astronomical Institute ``Anton Pannekoek,'' University of Amsterdam
and Center for High-Energy Astrophysics, Kruislaan 403, 1098 SJ
Amsterdam, The Netherlands}
\email{michiel@astro.uva.nl}

\author{Jeffrey E. McClintock and Michael R. Garcia}
\affil{Harvard-Smithsonian Center for Astrophysics, 60 Garden Street,
Cambridge, MA 02138}
\email{jem@cfa.harvard.edu,garcia@head-cfa.harvard.edu}

\author{Paul J. Callanan}
\affil{Department of Physics, University College, Cork, Ireland}
\email{paulc@ucc.ie}

\author{Charles D. Bailyn}
\affil{Yale University, Department of Astronomy, P.O. Box 208101,
New Haven, CT 06520-8101}
\email{bailyn@astro.yale.edu}

\author{Raj K. Jain}
\affil{Department of Physics, Yale University, P.O. Box 208120, New Haven,
CT 06520-8120}
\email{rjain@astro.yale.edu}

\and

\author{Ronald A. Remillard}               
\affil{Center for Space Research, Massachusetts Institute of Technology,
Cambridge, MA 02139-4307}
\email{rr@space.mit.edu}

\altaffiltext{1}{Based on observations collected at the 
European Southern Observatory, Chile (program 65.H-0360) and
the William Herschel Telescope 
operated on the island of La Palma by the Isaac Newton 
Group in the Spanish Observatorio del Roque de los
Muchachos of the Instituto de Astrofisica de Canarias}

\altaffiltext{2}{Also at Astronomical Institute, Utrecht University}

\begin{abstract}
Spectroscopic observations of the fast X-ray transient and
superluminal jet source SAX J1819.3-2525 (V4641 Sgr) reveal a best
fitting period of $P_{\rm spect}=2.81678\pm 0.00056$ days and a
semiamplitude of $K_2=211.0\pm 3.1$ km s$^{-1}$.  The optical mass
function is $f(M)=2.74\pm 0.12\,M_{\odot}$.  We find a photometric
period of $P_{\rm photo}=2.81730 \pm 0.00001$ days using a light curve
measured from photographic plates.  The folded light curve resembles
an ellipsoidal light curve with two maxima of roughly equal height and
two minima of unequal depth per orbital cycle.  The secondary star is
a late B-type star which has evolved off the main sequence.  Using a
moderate resolution spectrum ($R=7000$) we measure $T_{\rm eff}=10500
\pm 200$~K, $\log g=3.5\pm 0.1$, and $V_{\rm rot}\sin i=123\pm 4$ km
s$^{-1}$ (1$\sigma$ errors).
Assuming synchronous rotation, our
measured value of the projected rotational velocity implies a mass
ratio of $Q\equiv M_1/M_2=1.50\pm 0.08$ (1$\sigma$).  The lack of
X-ray eclipses implies an upper limit to the inclination of $i\le
70.7^{\circ}$.  On the other hand, the large amplitude of the folded
light curve ($\approx 0.5$ mag) implies a large inclination ($i\gtrsim
60^{\circ}$).  Using the above mass function, mass ratio, and
inclination range, the mass of the compact object is in the range
$8.73 \le M_1 \le 11.70\,M_{\odot}$ and the mass of the secondary star
is in the range $5.49 \le M_2\le 8.14\,M_{\odot}$ (90\% confidence).
The mass of the compact object is well above the maximum mass of a
stable neutron star and we conclude that V4641 Sgr contains a black
hole.  The B-star secondary is by far the most massive, the hottest,
and the most luminous secondary of the dynamically confirmed black
hole X-ray transients.  We find that the $\alpha$-process elements
nitrogen, oxygen, calcium, magnesium, and titanium may be overabundant
in the secondary star by factors of two to 10 times with respect to
the sun.  Finally, assuming $E(B-V)=0.32\pm 0.10$, we find a distance
$7.40 \le d \le 12.31$ kpc (90\% confidence).  This large distance and
the high proper motions observed for the radio counterpart make V4641
Sgr possibly the most superluminal galactic source known with an
apparent expansion velocity of $\gtrsim 9.5c$ and a bulk Lorentz
factor of $\Gamma\gtrsim 9.5$, assuming the jets were ejected during
during one of the bright X-ray flares observed with the
{\em Rossi X-ray Timing Explorer}.
\end{abstract}

\keywords{binaries: spectroscopic --- black hole physics ---
stars: individual (V4641 Sgr) --- X-rays: stars}

\section{Introduction}

SAX J1819.3-2525 was discovered as a relatively faint X-ray source
independently with the Wide Field Cameras on {\em BeppoSAX} on 1999
February 20 \citep{zan99,int00} and with the Proportional Counter
Array on the {\em Rossi X-ray Timing Explorer} on 1999 February 18
\citep{mar99}.  The source had two rapid and bright X-ray flares
around 1999 September 15.  The first flare had a peak intensity of
about 4.5 Crab in the 2-12 keV X-ray band \citep{smi99} and 5 Crab in
20-100 keV X-ray band \citep{mcc99}.  The second peak, which came
about 0.8 days later, reached intensities of 12.2 and 8 Crab in the
two bands, respectively.  A few days after the giant flares the source
was no longer detected in X-rays [e.g.\ \citet{wij00}].

SAX J1819.3-2525 was detected as a bright radio source (0.4 Jy) on
1999 September 16 \citep{hje00}, about 16 hours after the second X-ray
flare.  The radio source was marginally resolved in this first
observation, indicating high expansion velocities.  Thereafter, the
source was detected at radio wavelengths for about three weeks.
\citet{hje00} modelled the radio images and the detailed radio light
curves as a combination of the ejection of a relativistic, freely
expanding jet and a subsequent ejection of a more slowly decaying,
optically thin jet segment.  The radio observations did not resolve
the moving components of the jets, so as a result one must make an
assumption about when the jets were ejected in order infer the proper
motion.  If  the jets were ejected during  the
brief 4.5  Crab X-ray flare, the apparent proper motion is
0.22 arcseconds per day.  \citet{hje00} concluded the most likely time
of the jet ejection was during the rise of the 12 Crab X-ray flare, in
which case the apparent proper motion would be 0.36 arcseconds per day.  
These high proper motions obviously imply large velocities;
an apparent proper motion of $\gtrsim
0.22$ arcseconds per day corresponds to an apparent velocity of
$\gtrsim 1.28cd$, where $c$ is the velocity of light and $d$ is the
distance in kpc.

SAX J1819.3-2525 was already known as an optical variable star
\citep{gor78}, which was discovered 21 years before the first known
X-ray activity.  The source had a $\approx 2$ mag flare in 1978 June,
and otherwise showed variability with an amplitude of up to $\approx
1$ mag about a mean level of $B\approx 14.2$ \citep{gor90}.  In spite
of differences in the classifications and variability behavior,
Goranskij's variable was confused with GM Sgr, a long-period variable
star discovered by \citet{luy27}.  As a result, many of the papers
discussing the optical counterpart of SAX J1819.3-2525 refer to GM
Sgr.  \citet{haz99} located the plate material used by Luyten and
found that Luyten's variable GM Sgr is actually one arcminute south of
Goranskij's variable star, which is the star at the precise radio
coordinates of the X-ray transient.  Goranskij's variable has been
given the new variable star designation V4641 Sgr, and we shall use
this name for the remainder of the paper.  \citet{gor90} assembled a
quiescent light curve using 345 plates taken over a span of about 30
years with the Crimean 40 cm astrograph.  He noted a periodicity at a
frequency of 2.7151 cycles per day, corresponding to a double-wave
period of 0.7365483 days.  However, owing to severe aliasing problems,
\citet{gor90} could not rule out 1 day aliases of his preferred
period.  We demonstrate below that the orbital period is 2.8713 days,
which is an alias of the period noted above (i.e.\
$1/(0.5*2.8173)\approx 0.71$ cycles per day).

A network of optical observers had noted an increase in the
variability of V4641 Sgr just prior to the giant X-ray flares
\citep{kat99}.  About six days before these flares the source showed a
$\approx 1$ mag modulation with a period of $\approx 2.5$ days.  The
optical brightness peaked at $V=8.8$ on 1999 September 15.8, and
subsequently decayed rapidly to its mean quiescent level within two
days.  An optical spectrum obtained 1999 September 16.25 showed very
strong Balmer emission lines \citep{djo99}.  The Balmer lines made an
abrupt transition from emission to absorption between 1999 September
17 and 19 \citep{gar99}.  A spectrum obtained 1999 September 30
\citep{wag99} showed only strong Balmer absorption lines and
interstellar features.  At the time, this was quite confusing since
the quiescent counterpart was thought to be an early K-type star based
on the spectrum of GM Sgr shown in \citet{dow95}.  However, in
hindsight, the Balmer absorption line spectrum makes sense since we
now know that \citet{dow95} did not observe V4641 Sgr, which in fact
contains a late B star.  It now appears that the quiescent optical
state was reached by 1999 September 19 (i.e.\ about four days after
the X-rays were undetectable).

It was initially thought that V4641 Sgr was relatively nearby based on
two arguments.  First, if the intrinsic velocity of the jet(s)
observed in V4641 Sgr is similar to the velocities observed in GRO
J1655-40 and GRS 1915+105 \citep{mir99}, then the large observed
proper motion of the jet observed in V4641 Sgr would imply $d\approx
500$ pc.  Second, a K-star in a $\approx 0.7$ day orbit with an
apparent $V$ magnitude of $\approx 13.7$ would have a distance of
$\approx 1$ kpc.  We demonstrate below that $d\ge 7.4$ kpc at the 90\%
confidence level.  Thus the apparent velocity of the 0\farcs 25 jet(s)
seen several hours after the beginning of the X-ray event was $\gtrsim
9.5c$, assuming the jets were ejected at the start of the giant
X-ray flare.

The galactic ``microquasars'' are excellent laboratories for the study
of relativistic jets since they evolve orders of magnitude more
quickly than the jets in quasars evolve \citep{mir99}.  V4641 Sgr is
of special interest because of its very rapid X-ray flaring behavior and
its extreme superluminal jets, and because the source is optically
bright in quiescence
($V\approx 13.7$).  In this paper we report the results of our
spectroscopic observations of V4641 Sgr.  The observations and basic
data reductions are summarized in Section \ref{obssec}.  In Section
\ref{persec} we establish the basic orbital parameters of the system.
The modelling of a high resolution spectrum and the derivation of the
secondary star properties are discussed in Section \ref{secsec}.  We
discuss the astrophysical parameters of the binary system in Section
\ref{impsec} and discuss the implications of our results in Section
\ref{dissec}.  The paper ends with a short summary in Section
\ref{sumsec}.

\section{Observations}\label{obssec}

We obtained a total of 45 spectra of V4641 Sgr between 1999 September
17 and 1999 October 16 using the FAST spectrograph \citep{fab98} on
the 1.5 m telescope at the Fred L. Whipple Observatory (FLWO) on Mount
Hopkins, Arizona.  The 300 g mm$^{-1}$ grating was used with a Loral
$512\times 2688$ CCD.  About half of the observations were taken with
a 1\farcs 5 wide slit, yielding a spectral resolution of $\approx
4.1$~\AA\ FWHM.  The other observations were taken with a 5\farcs 0
slit, yielding a spectral resolution of $\approx 6.0$~\AA\ FWHM.  The
spectra cover the wavelength range 3650-6800~\AA.  The exposure times
were from three to ten minutes and the spectra were generally obtained
under mostly clear skies with seeing between 1 and 2 arcseconds.  The
spectrograph slit was rotated to maintain an approximate alignment
with the parallactic angle, and a wavelength calibration lamp was
observed after each exposure.

As we noted above, the source was active until about 1999 September
18.  All of the spectra from September 17 and 18 show a broad
H$\alpha$ emission line, with a full width at zero intensity of up to
$\approx 5000$ km s$^{-1}$.  In contrast, the higher Balmer lines were
always in absorption.  By September 19, the H$\alpha$ line was also in
absorption, and apparently the source was near or at its quiescent
level.  In the present paper we will discuss 14 spectra obtained in
quiescence (i.e.\ between 19 September and 16 October) that were taken
through the 1\farcs 5 wide slit.

A total of forty-seven additional spectra of the source were obtained
2000 June 4-7 using the FORS1 instrument on Antu, which is the first
8.2 m telescope at the European Southern Observatory, Paranal.
Thirty-seven of the spectra were taken with the 600B grism and a
0\farcs 7 wide slit; this combination gives a spectral resolution of
4~\AA\ FWHM and a wavelength coverage of 3366-5751~\AA.  The other ten
spectra were taken with the 600R grism and a 0\farcs 7 wide slit,
yielding a spectral resolution of 3.2~\AA\ FWHM and a wavelength
coverage of 5142-7283~\AA.  The seeing was relatively poor for the VLT
($\approx 1\farcs 5$), and clouds were present on the night of June
5-6.  The exposure times were from 1.5 to 4 minutes depending on the
conditions.  An atmospheric dispersion corrector was used, so the slit
was kept at the default north-south direction.  Following the standard
procedure at Paranal, the flat-field and wavelength calibration
exposures were obtained during the daytime hours with the telescope
pointed at the zenith.

Ten additional spectra of the source were obtained 2000 June 16 and 17
with the 3.5 m New Technology Telescope (NTT) at the European Southern
Observatory, La Silla, using the blue arm of EMMI.  The instrumental
configuration consisted of the Tek $1024\times 1024$ CCD, a 1\farcs 0
wide slit, and the grating \#11, which is a high efficiency
holographic grating with 3000 ``grooves'' per mm \citep{wil91}.  The
spectral resolution is 0.43~\AA\ FWHM and the wavelength coverage is
3870.6-4032.8~\AA.  The slit was aligned with the parallactic angle,
and a thorium-argon lamp was observed for wavelength calibration.  The
weather was clear June 16 and a total of eight 30-minute exposures
were obtained.  The conditions on June 17 were quite poor, and only
two 30-minute exposures were taken.

Finally, seven additional spectra were obtained 2000 August 12 and 13
with the 4.2 m William Herschel Telescope (WHT) at the Observatorio de
Roque de los Muchachos, la Palma, Spain.  The instrumental
configuration was the blue arm of the ISIS double beam spectrograph,
the R1200B grating, and the EEV12 $4096\times 2048$ CCD.  We used a
0\farcs 8 wide slit, yielding a spectral resolution of 0.58~\AA\ FWHM
and a coverage of 3680-4500~\AA.  The slit was aligned with the
parallactic angle, and a copper-argon lamp was observed for wavelength
calibration.  The weather was photometric, and the seeing was
generally 1\farcs 1 or less.

We reduced all of the spectra using IRAF\footnote{IRAF is distributed
by the National Optical Astronomy Observatories, which are operated by
the Association of Universities for Research in Astronomy, Inc., under
cooperative agreement with the National Science Foundation.}.  The
standard tasks were used to apply the bias and flat-field corrections,
and to extract and wavelength-calibrate the spectra.  The bright night
sky emission line at $\approx 5578$~\AA\ was used to make small
adjustments to the wavelength scales of the FLWO and VLT spectra.  The
shifts required to align this feature to a common wavelength
(5578.0~\AA) were generally less than $\approx 15$ km s$^{-1}$,
although the VLT spectra from the end of the night of June 7 required
shifts on the order of 80 km s$^{-1}$.

\begin{figure*}[t]
\epsscale{1.0}
\plotone{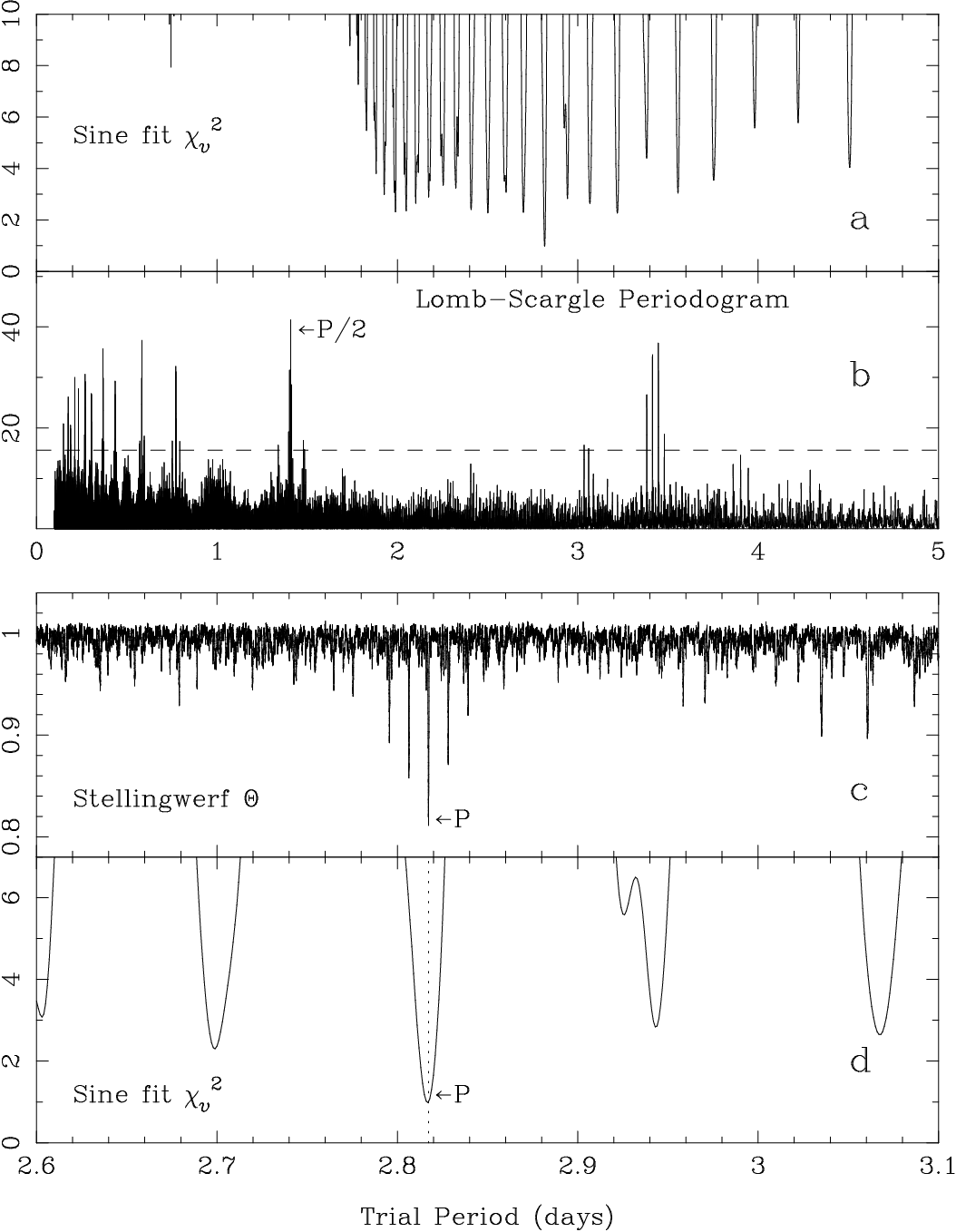}
\figcaption[f1.ps]{(a): The solid line shows the reduced
$\chi^2_{\nu}$ for a three-parameter sinusoid fit to the 71 radial
velocities as a function of the trial period in days.  The minimum
$\chi^2_{\nu}$ is at $P=2.8177\pm 0.0005$ days.  (b): The solid line
shows the Lomb-Scargle periodogram of the photometric light curve.
Peaks above the dashed line are significant at the $3\sigma$ level.
The peak with the most power is at $P=1.40865$ days, which is half of
the spectroscopic period.  (c): The Stellingwerf (1978) $\Theta$
statistic as a function of the trial period near 2.8 days.  The
minimum $\Theta$ and hence the most coherent modulation is at
$P=2.81730\pm 0.00001$ days.  (d): An expanded view of the curve in
(a).  The various aliases of the spectroscopic period are ruled out by
the photometry.
\label{fig1}}
\end{figure*}

\section{Orbital Period and Spectroscopic Elements}\label{persec}

\subsection{Spectroscopic Period}

We measured the radial velocities of the secondary star using the {\em
fxcor} task within IRAF, which is an implementation of the technique
of \citet{ton79}.  Synthetic spectra with $T_{\rm eff}=10500$, $\log
g=3.5$, and $V_{\rm rot}\sin i=123$ km s$^{-1}$ (see below) were used
as templates for the five different sets of spectra (i.e.\ FLWO, VLT
``blue'', VLT ``red'', NTT, and WHT).  The cross correlation regions
covered the available wavelengths between 3700 and 6847~\AA, excluding
the interstellar Ca II H and K and Na D lines, the diffuse
interstellar bands near 4330 and 5780~\AA, and a telluric feature near
6280~\AA.  The final velocities for the FLWO, the VLT blue, and WHT
spectra were insensitive to the exact starting and ending points of
the cross correlation regions and to the function used to find the
centroid of the cross correlation peak.  On the other hand, the
velocities of the VLT red spectra and the NTT spectra were sensitive
to these parameters since these spectra contain very few lines.  We
can estimate approximate velocity corrections for the VLT red and NTT
spectra by using observations of an A0 template star observed both
with the VLT and NTT.  However, we find that including the corrected
velocities did not improve the resulting spectroscopic parameters
since the corrected velocities all have relatively large
uncertainties.  We therefore excluded the VLT red and NTT velocities
from the analysis presented below.

\begin{figure*}[t]
\epsscale{1.0}
\plotone{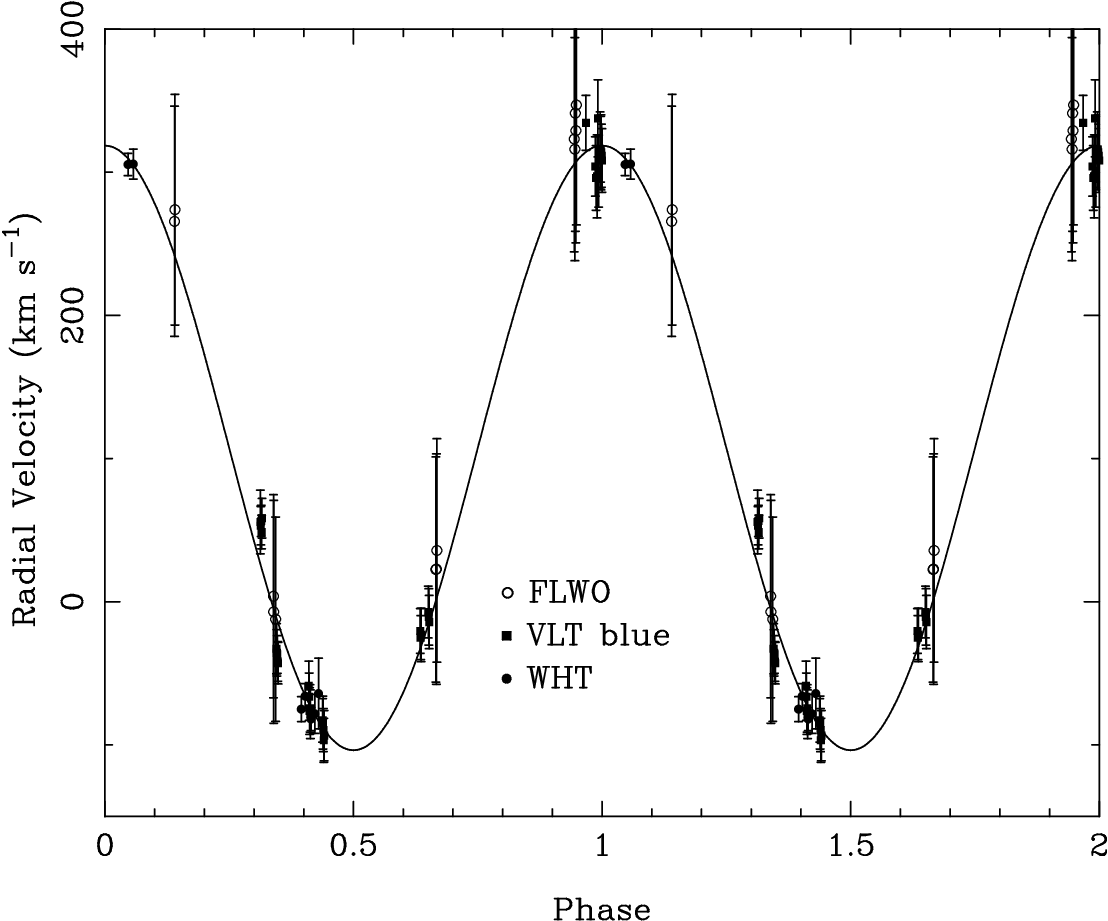}
\figcaption[f2.ps]{The radial velocities for V4641 Sgr folded on
the spectroscopic period and phase and the best fitting sinusoid.  See
Table \protect\ref{tab2} for the parameters.  Each point has been
plotted twice for clarity.
\label{fig2}}
\end{figure*}

To search for the spectroscopic period we computed a three-parameter
sinusoid fit to the velocities for a range of trial periods between 0
and 5 days and recorded the values of the reduced $\chi^2$ for the
fits.  The free parameters at each trial period are the velocity
semiamplitude $K_2$, the epoch of maximum velocity $T_0$(spect), and
the systemic velocity $\gamma$.  The error bars on the individual
velocities were scaled by a factor of 2.65 to yield $\chi^2_{\nu}=1.0$
at the minimum.  Fig.\ \ref{fig1}a shows the resulting $\chi^2_{\nu}$
vs.\ $P$ curve.  The best fit is at a period of $P=2.81678\pm 0.00056$
days ($1\sigma$ error).  The next best fit occurs at $P=2.699$ days,
where $\chi^2_{\nu}=2.27$.  This alias period and the others are
clearly ruled out by inspection of the folded velocity curves and we
adopt the following spectroscopic elements: $P_{\rm spect}=2.81678\pm
0.00056$ days, $K_2=211.0\pm 3.1$ km s$^{-1}$, $\gamma=107.4\pm 2.9$
km s$^{-1}$, 
and $T_0({\rm
spect})={\rm HJD~}2,451,442.523\pm 0.052$.  The resulting optical mass
function is then $f(M)=2.74\pm 0.12\,M_{\odot}$ ($1\sigma$ errors).
Fig.\ \ref{fig2} shows the velocities and the best fitting sinusoid,
and the spectroscopic elements are listed in Table \ref{tab2}.

\subsection{Photometric Period}

\citet{gor90} published a light curve of V4641 Sgr obtained from 345
plates taken with the Crimean 0.40 m astrograph between about 1960 and
1990.  After it was realized that V4641 Sgr was the optical
counterpart of SAX J1819.2-2525, Goranskij reanalyzed the photographic
data and made them publically available via the
VSNet\footnote{http://www.kusastro.kyoto-u.ac.jp/vsnet/Mail/obs26000/msg00925.html}.
We searched these data for periodicities using both the technique of
\citet{ste78} and the Lomb-Scargle Periodogram \citep{lom76,sca82}.
We excluded the uncertain measurements and also data from 1978 June,
when the source had an optical flare.  Fig.\ \ref{fig1}b shows the
Lomb-Scargle periodogram in the period range of 0 to 5 days.  Using a
Monte Carlo procedure, we computed the power above which the peaks are
significant at the $3\sigma$ level.  We made 10,000 simulated light
curves having the same observation times as the actual light curve and
with magnitude points normally distributed about the mean
magnitude. (By construction, the simulated light curves have the same
variance as the actual light curve.)  A Lomb-Scargle periodogram was
computed for each simulated light curve and the power of the strongest
peak was recorded.  The $3\sigma$ threshold was then determined from
the cumulative distribution of the peak powers.  This threshold is
shown by the dashed line in Fig.\ \ref{fig1}b.  There are several
peaks significant at the $3\sigma$ level.  The peak with the most
power is at a trial period of $P=1.40865$ days (0.7099 cycles
day$^{-1}$), which is half of the spectroscopic period.  All of the
other significant peaks are aliases of the main peak (e.g.\
$1/1-1/1.40865=1/3.447$, etc.) and their sidebands.  Fig.\ \ref{fig1}c
shows the $\Theta$ statistic of \citet{ste78} plotted as a function of
the trial period in the range $2.6\le P\le 3.1$ days (we used 20 phase
bins of width 0.05 to compute the $\Theta$ statistic---this
combination yielded the smoothest minima).  The minimum $\Theta$ is at
a period of $P=2.81730$ days.  Yearly aliases of this period are also
evident.  Fig.\ \ref{fig1}d shows the $\chi^2_{\nu}$ vs.\ $P$ curve
from Fig.\ \ref{fig1}a plotted over the same period range as in Fig.\
\ref{fig1}c.  The period of $P=2.81730$ days is the only trial period
where there are minima in both the $\Theta$ statistic and in the
$\chi^2_{\nu}$ vs.\ $P$ curve.  Thus the photometric and spectroscopic
periods of V4641 Sgr are nicely consistent with each other.

\begin{figure*}
\epsscale{1.0}
\plotone{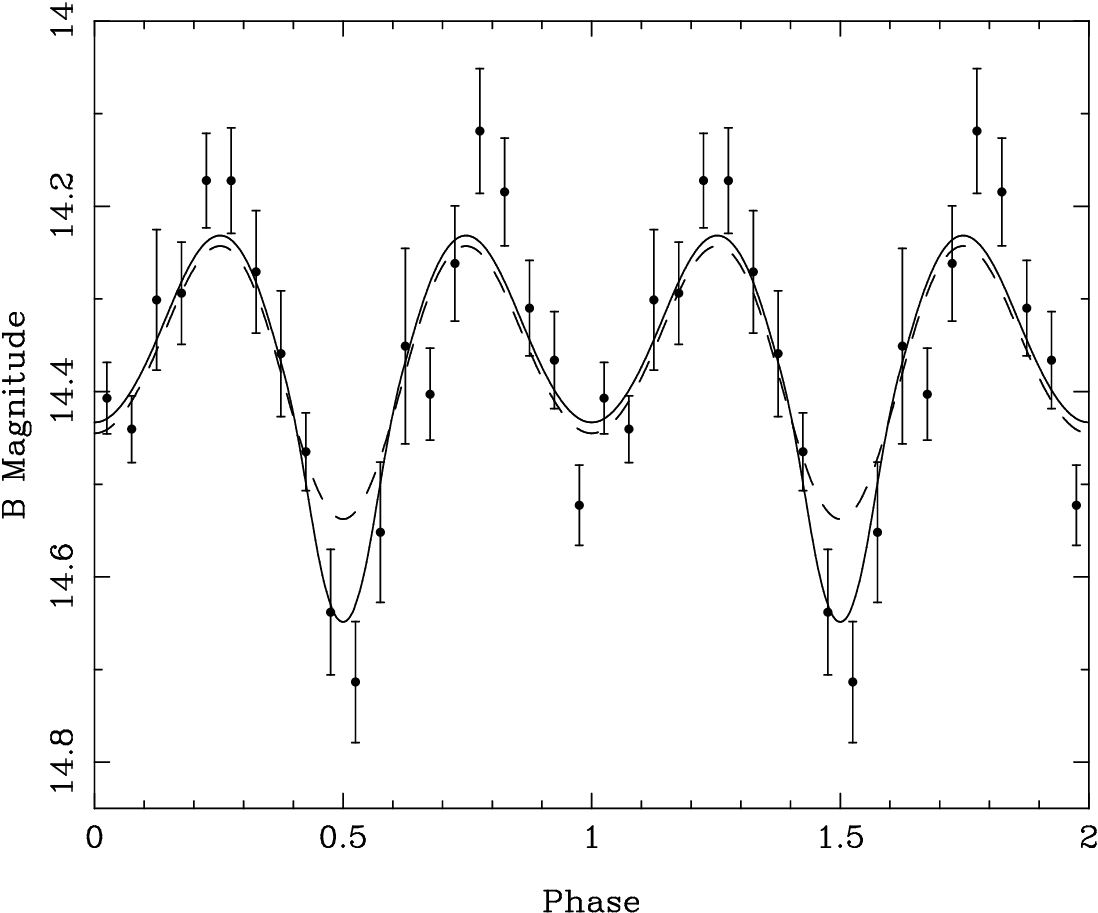}
\figcaption[f3.ps]{The photometric light curve of V4641 Sgr
folded using $P=2.81730$ days and $T_0=T_0$(photo)$+P/2$ and binned
into 20 bins of equal size.  Each point has been plotted twice for
clarity.
The error bars on the points represent
the error of the mean. 
The solid
curve is a $B$-band ellipsoidal model
with $i=70^{\circ}$, $Q=1.6$, and a grazing eclipse of the star by an
accretion disk.  The dash curve is the resulting light curve when the
disk is removed.
\label{fig3}}
\end{figure*}

We computed the uncertainty in the photometric period by using another
Monte Carlo procedure.  We adopted a mean photometric error of 0.1 mag
(Goranskij, private communication) and constructed 10,000 simulated
data sets by assigning a new magnitude to each time using $y_{{\rm
new},i}=y_i+0.1\sigma$, where $\sigma$ is a random Gaussian deviate
with variance of unity and a zero mean.  For each simulated data set
we computed the Lomb-Scargle periodogram and the Stellingwerf $\Theta$
statistic and recorded the resulting derived periods.  The uncertainty
in the photometric period was taken to be the standard deviation of
the distribution of best periods from these simulated data.  We find
$P=2.81730\pm 0.00001$ days.  Fig.\ \ref{fig3} shows the light curve
folded on this period and binned into 20 phase bins. The light curve
resembles an ellipsoidal light curve with two maxima of roughly equal
heights and two minima of unequal depths per orbital cycle.  The
amplitude is quite large; the range is about 0.5 mag from the deeper
minimum to the maximum.

\citet{gor90} gives the time of minimum light $T_0$(photo) as HJD
2,447,707.454 (no error is quoted).  We measured $T_0$(photo) by
folding the light curve on a dense grid of trial values of
$T_0$(photo) and recording the $\chi^2$ of an ellipsoidal model fit
(the solid line in Fig.\ \ref{fig3}).  The result is $T_0$(photo) =
HJD 2,447,707.$4865\pm 0.0038$.  The photometric phase of the time of
maximum velocity $T_0$(spect) is $0.75\pm 0.02$, exactly as expected
for an ellipsoidal light curve. (Note that the phase zero point used
in Fig.\ \ref{fig3} is $T_0$(photo)$+0.5P$ to conform with our
ellipsoidal modelling code.)

\section{Parameters for the Secondary Star}\label{secsec}

\begin{figure*}[t]
\epsscale{1.0}
\plotone{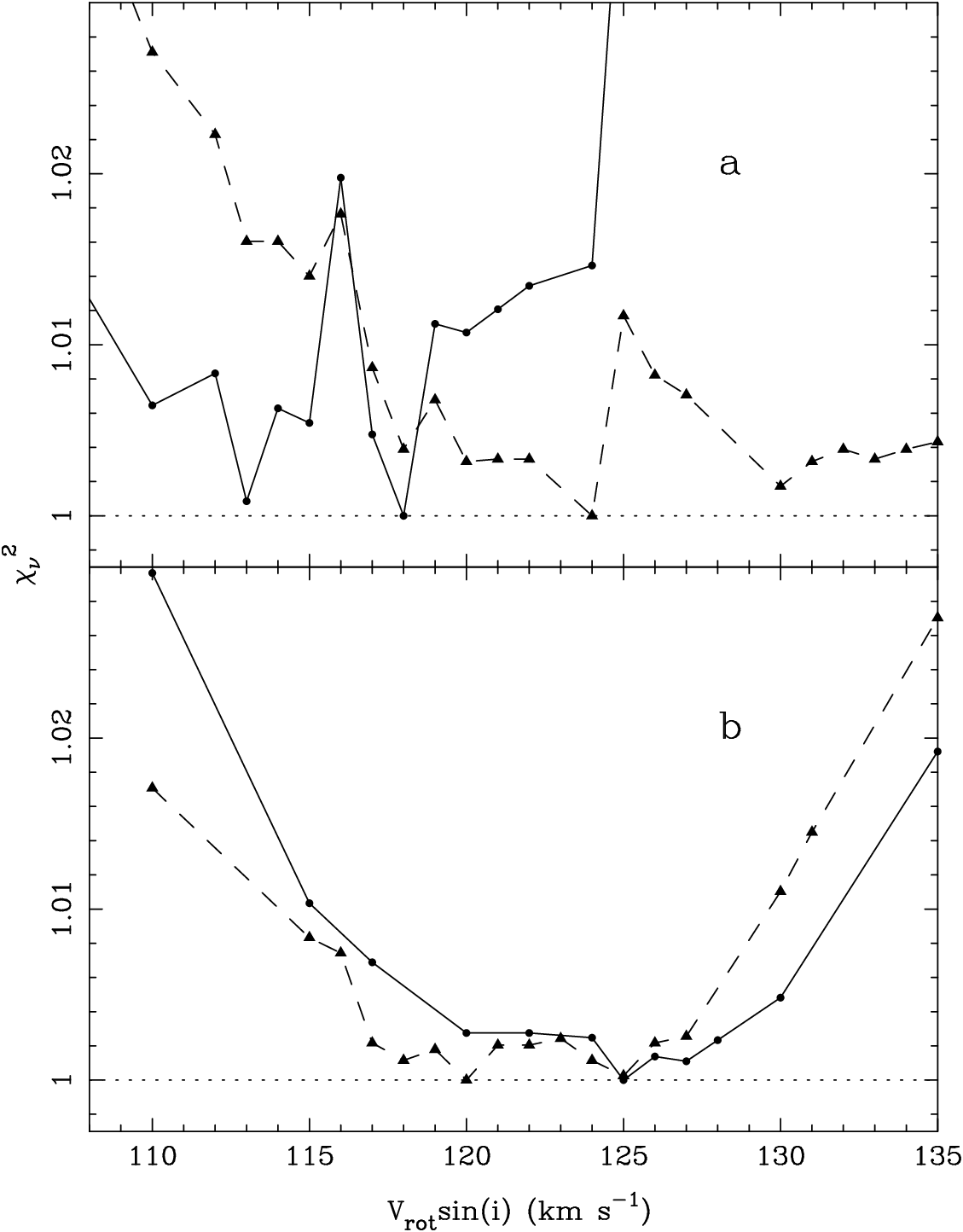}
\figcaption[f4.ps]{(a):  The reduced 
$\chi^2$
as a function of the input
value of $V_{\rm rot}\sin i$ for the August 13 WHT restframe spectrum
(solid line) and August 12 WHT restframe spectrum (dashed line), computed
using the model spectra broadened by SYNPLOT.  (b):
The reduced 
$\chi^2$
as a function of the input
value of $V_{\rm rot}\sin i$ for the August 13 WHT restframe spectrum
(solid line) and August 12 WHT restframe spectrum (dashed line), computed
using the numerical broadening kernels.
\label{rmsfig}}
\end{figure*}

\begin{figure*}[t]
\epsscale{1.0}
\plotone{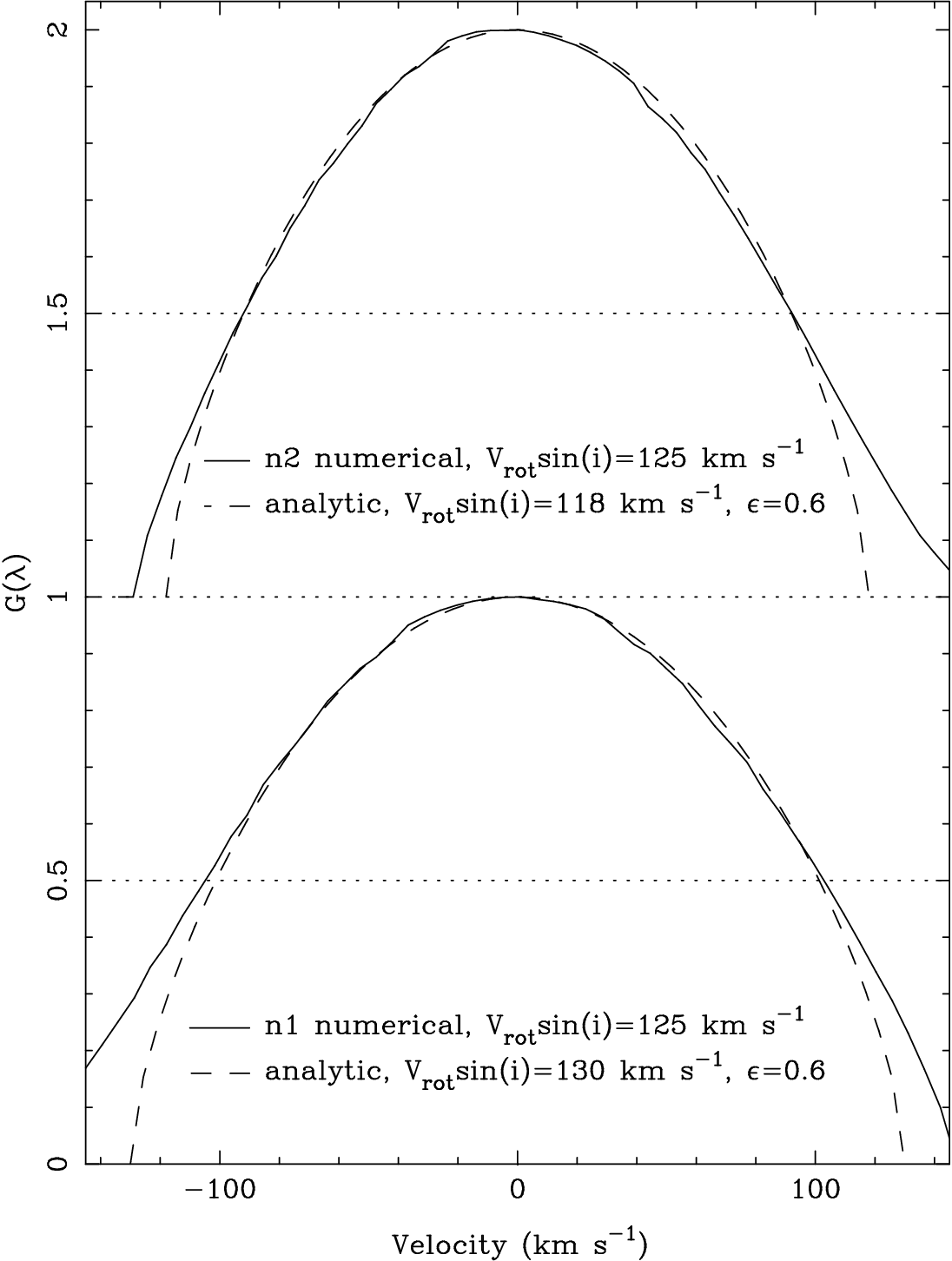}
\figcaption[f5.ps]{Top:
The solid line is a numerical rotational 
broadening kernel
which is the mean of 5 separate kernels matched to the August 13 WHT
spectra, scaled by 125 km s$^{-1}$.
The dashed line is the analytic kernel used by {\sc SYNPLOT}, scaled by
118 km s$^{-1}$.
Bottom: 
The solid line is a numerical rotational 
broadening kernel
which is the mean of the two separate kernels matched to the August 12 WHT
spectra, scaled by 130 km s$^{-1}$.
The dashed line is the analytic kernel used by {\sc SYNPLOT}, scaled by
118 km s$^{-1}$.  
\label{fig5}}
\end{figure*}

V4641 Sgr is relatively bright, and as a result one can easily obtain
a spectrum with a relatively high signal-to-noise ratio (SNR) and high
resolution.  The eight NTT spectra from June 16 were Doppler shifted
to zero velocity and combined to yield a ``restframe'' spectrum with a
resolution of $R=9000$ and a SNR of about 50 per pixel.  In the same
way, we also made a restframe spectrum using the WHT data.  The SNR of
the WHT restframe spectrum is about 130 per pixel and is reasonably
constant over much of the spectrum.  We used synthetic spectra
generated from solar metallicity Kurucz
models\footnote{http://cfaku5.harvard.edu/} to determine three basic
parameters for the secondary star, namely its effective temperature
$T_{\rm eff}$, surface gravity $\log g$, and mean projected rotational
velocity $V_{\rm rot}\sin i$.  The {\sc IDL} program {\sc SYNPLOT}
\citep{hub94} was used to compute the detailed model spectra.  {\sc
SYNPLOT} can generate an optical spectrum with arbitrary spectral
resolution, wavelength sampling, and rotational broadening. [Currently
{\sc SYNPLOT} uses an analytic broadening kernel with a linear limb
darkening coefficient of $\epsilon=0.6$ \citep{gra92}].  For the NTT
spectrum the resolution was taken to be $0.4333$~\AA\ and the
wavelength sampling was 0.1530~\AA, and for the WHT spectra those
quantities were 0.5802~\AA\ and 0.2231~\AA, respectively.
We also have two synthetic spectra kindly provided by Peter Hauschildt
computed using his {\sc PHOENIX} code \citep{hau99}.  These models are
fully line-blanketed with magnesium and iron computed in non-LTE (NLTE).

We used the technique outlined in \citet{mar94} to fit the model
spectra to the observed ones.  This fitting procedure has the
advantage that it allows for the possibility of a (continuum)
contribution from the accretion disk.  The restframe spectra were
normalized to their continuum fits.  An input model spectrum
characterized by the three parameters $T_{\rm eff}$, $\log g$, and
$V_{\rm rot}\sin i$ is normalized to its continuum using the same
procedure used for the observed spectra: it is scaled by a weight
factor $w$, and subtracted from the observed spectrum.  The scatter in
the difference spectrum is measured by computing the reduced $\chi^2$
from a low-order polynomial fit (after the interstellar lines were
masked out of the fit).  We varied $T_{\rm eff}$, $\log g$, and
$V_{\rm rot}\sin i$ and looked for the ``smoothest'' difference
spectrum.  The Kurucz grid has models for gravities of $\log g=3.0$,
3.5, and 4.0 (cgs units) near the temperatures of interest (about
10000~K).

\subsection{Effective Temperature and Spectral Classification}

It quickly became clear using the NTT spectrum that the gravity of the
secondary star in V4641 Sgr is quite close to $\log g=3.5$ since the
$\log g=3.0$ models have Balmer lines that are much too narrow and the
$\log g=4.0$ models have Balmer lines that are much too broad.  We
therefore fixed the gravity at $\log g=3.5$.  The temperature is best
constrained by the WHT spectrum.  The He I line at 4024~\AA\ is
somewhat sensitive to the temperature, and we find $T_{\rm
eff}=10500$~K.  The statistical errors in the temperature and gravity
are much smaller than the respective grid spacings (0.5 dex in $\log
g$ and 250 K in $T_{\rm eff}$).  No doubt there are systematic errors
as well, but it is beyond our present ability to quantify these
errors.  We conservatively adopt $\sigma_T=200$~K and $\sigma_{\log
g}=0.1$.  The values of $T_{\rm eff}$ and $\log g$ we find for V4641
Sgr are not too different from those of the B9 III MK spectral type
standard $\gamma$ Lyr for which \citet{kun97} list $T_{\rm
eff}=9970\pm 540$ and $\log g=3.50$; the gravity was estimated from a
measurement of the H$\beta$ index \citep{bal86}.  We fit synthetic
spectra to the observed spectrum of $\gamma$ Lyr (obtained with the
WHT) and find $T_{\rm eff}=10000$~K, in good agreement with the
previous values, and $\log g=3.0$, somewhat lower than the previous
values.  Based on this approximate similarity, we assign a B9III
classification to the secondary star in V4641 Sgr.  
We note that the spectral type of B9III derived here
is somewhat earlier than the
preliminary type of A2V we announced in IAU Circular 7440.   The latter
value was arrived at via a visual inspection of the Balmer lines in
a low resolution spectrum.  We believe the spectral type of B9III derived
here is much more reliable since we used an impartial fitting procedure on
a spectrum with higher resolution and a higher SNR.

\begin{figure*}[t]
\epsscale{1.0}
\plotone{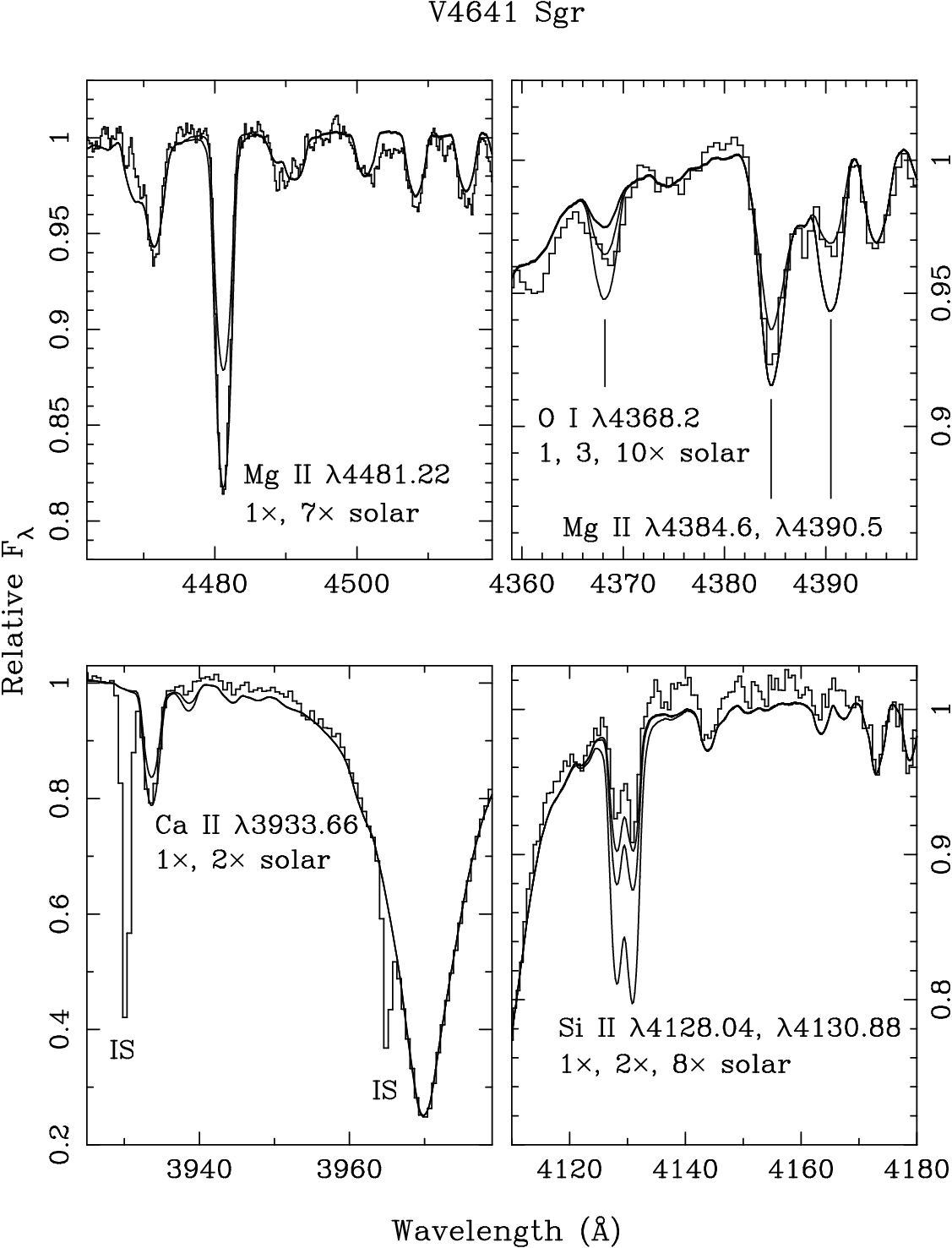}
\figcaption[f6.ps]{The results of our exploratory abundance analysis
for V4641 Sgr.
We show the solar abundance base Kurucz model, and the models with the
abundance patterns altered to approximately match the line profiles.
The observed spectrum is from the WHT
(resolution 0.58~\AA\ FWHM).  The spectrum
was rebinned by a factor of 2.6 for comparison
with the Ca II line and the O I line.  The interstallar Ca II lines are
labelled with ``IS''.
\label{plotabund}}
\end{figure*}

\begin{figure*}[t]
\epsscale{1.0}
\plotone{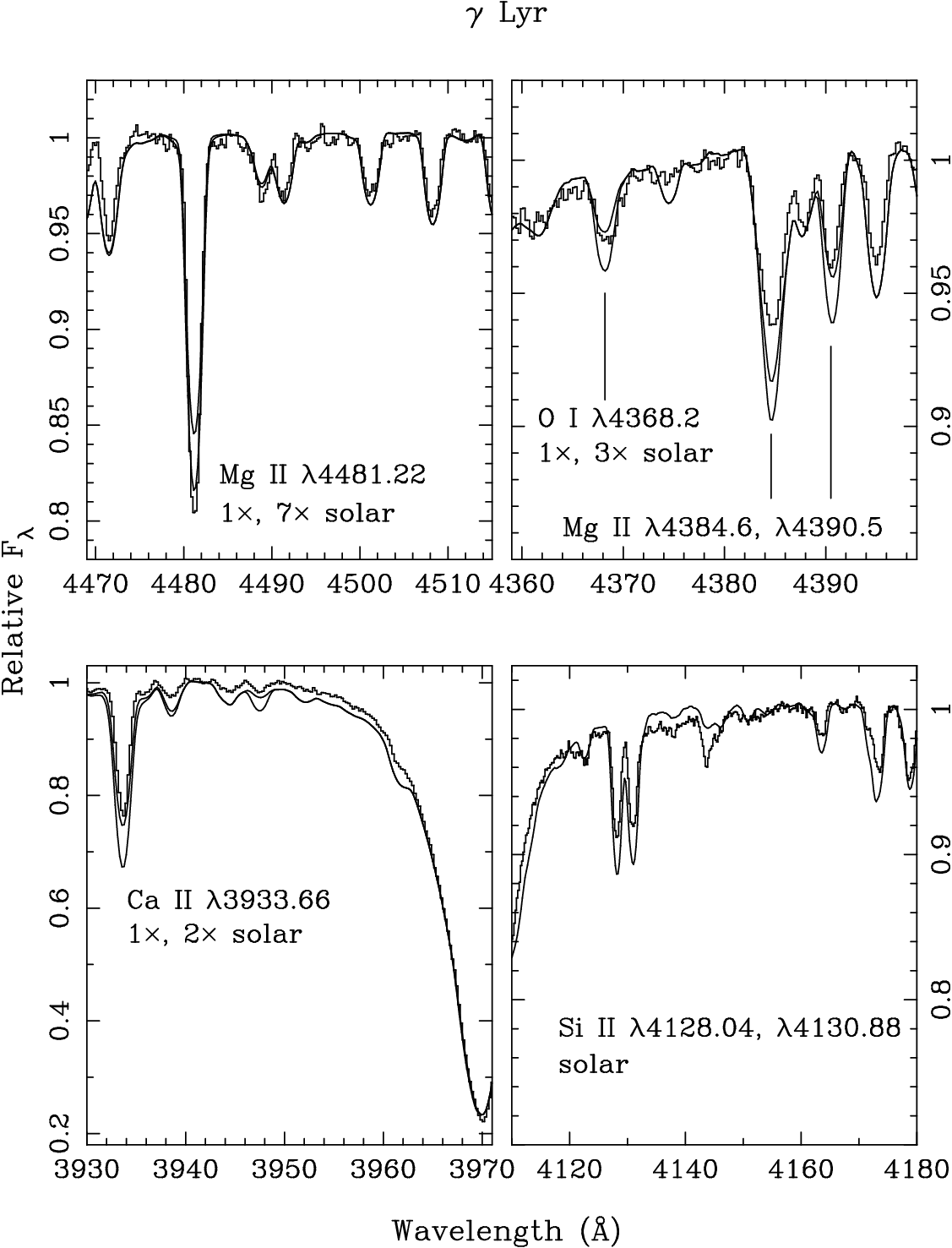}
\figcaption[f7.ps]{The abundance analysis
for $\gamma$ Lyr.
We show the solar abundance base Kurucz model
($T_{\rm eff}=10000$~K, $\log g=3.0$), and
$V_{\rm rot}\sin i=88$ km s$^{-1}$, and the models with the
abundance patterns found from the V4641 Sgr spectrum.  The observed
spectrum is from the WHT (resolution 0.58~\AA\ FWHM). 
\label{plotabundlyr}}
\end{figure*}

\begin{figure*}[t]
\epsscale{1.0}
\plotone{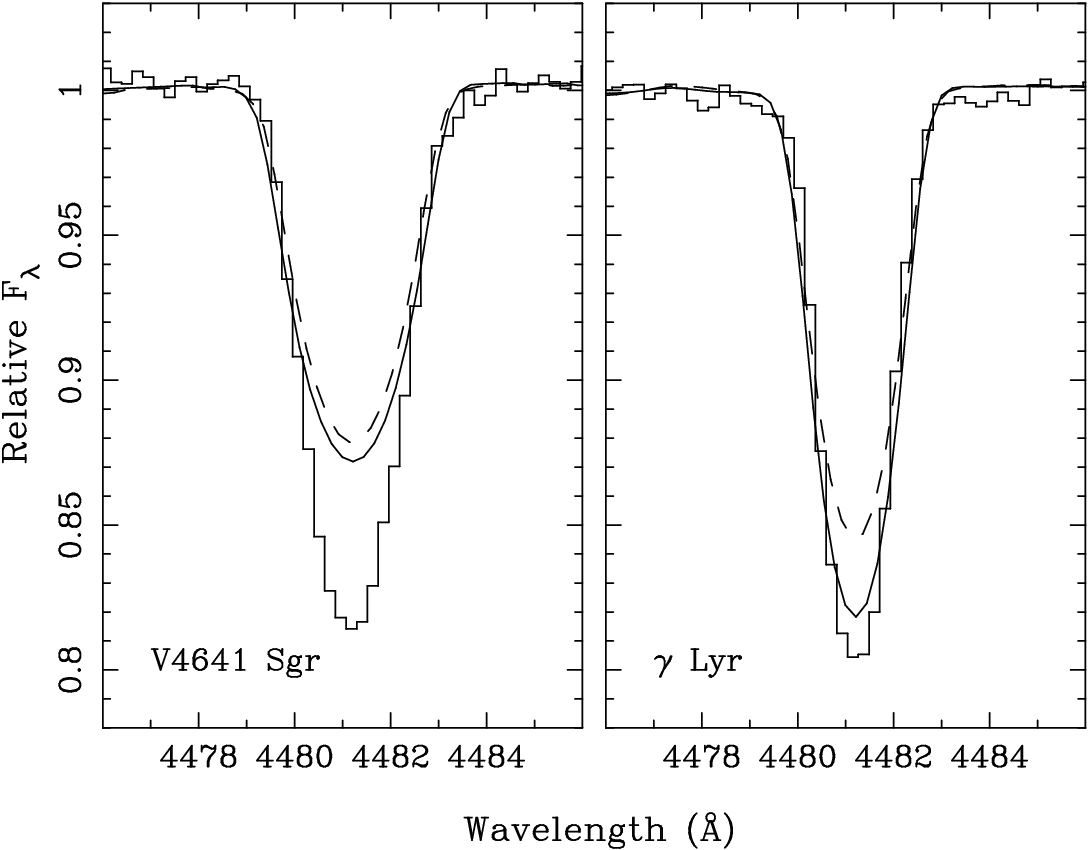}
\figcaption[f8.ps]{The Mg II $\lambda$4481.22 lines in V4641 Sgr
(left) and $\gamma$ Lyr (right), observed
with the WHT (resolution 0.58~\AA\ FWHM). 
The synthetic spectra made from
the LTE solar metallicity Kurucz models are shown as the dashed lines
and the Hauschildt model spectra (computed with magnesium in NLTE)
are shown as the
solid lines.  The  NLTE treatment of magnesium in $\gamma$ Lyr changes
the strength of the line significantly, whereas there is little change
in the line strength in V4641 Sgr. 
\label{nltemg}}
\end{figure*}

\begin{figure*}[t]
\epsscale{1.0}
\plotone{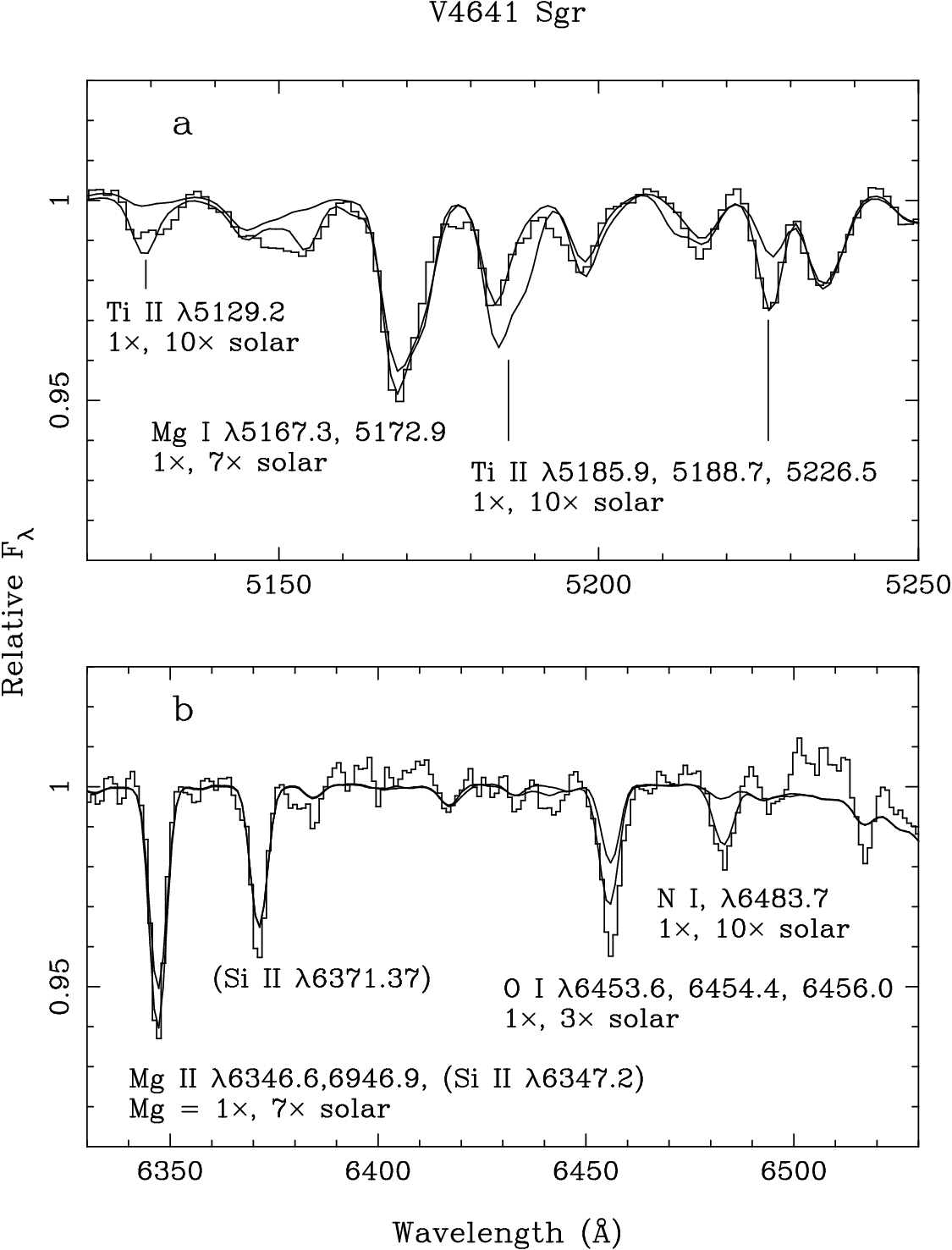}
\figcaption[f9.ps]{(a):  The restframe spectrum of V4641 Sgr
observed with the VLT FORS1 and
grism 600B (spectral resolution 4~\AA\ FWHM), the
solar abundance model
($T_{\rm eff}=10500$~K, $\log g=3.5$, and
$V_{\rm rot}\sin i=123$ km s$^{-1}$), and the Mg and Ti enhanced
models.
b:  The restframe spectrum of V4641 Sgr observed with the VLT FORS1
and grism 600R (spectral resolution 3.2~\AA\ FWHM),
the
solar abundance model
($T_{\rm eff}=10500$~K, $\log g=3.5$, and
$V_{\rm rot}\sin i=123$ km s$^{-1}$), and the N, O, and Mg enhanced
models
\label{abundlow}}
\end{figure*}

\subsection{Rotational Velocity of the Secondary}

Two other interesting facts also came to light in this process.
First, it was apparent that the rotational velocity of V4641 Sgr is
quite high ($\gtrsim 0.5K_2$) since the atomic lines are quite
broadened by rotation.  Second, some of the metal lines (most noteably
Mg II near 4781~\AA) are much stronger in V4641 Sgr than they are in
the models.  We therefore iterated the fits to determine the
rotational velocity and the approximate abundances of Mg and other
metals.

\begin{figure*}[t]
\epsscale{1.0}
\plotone{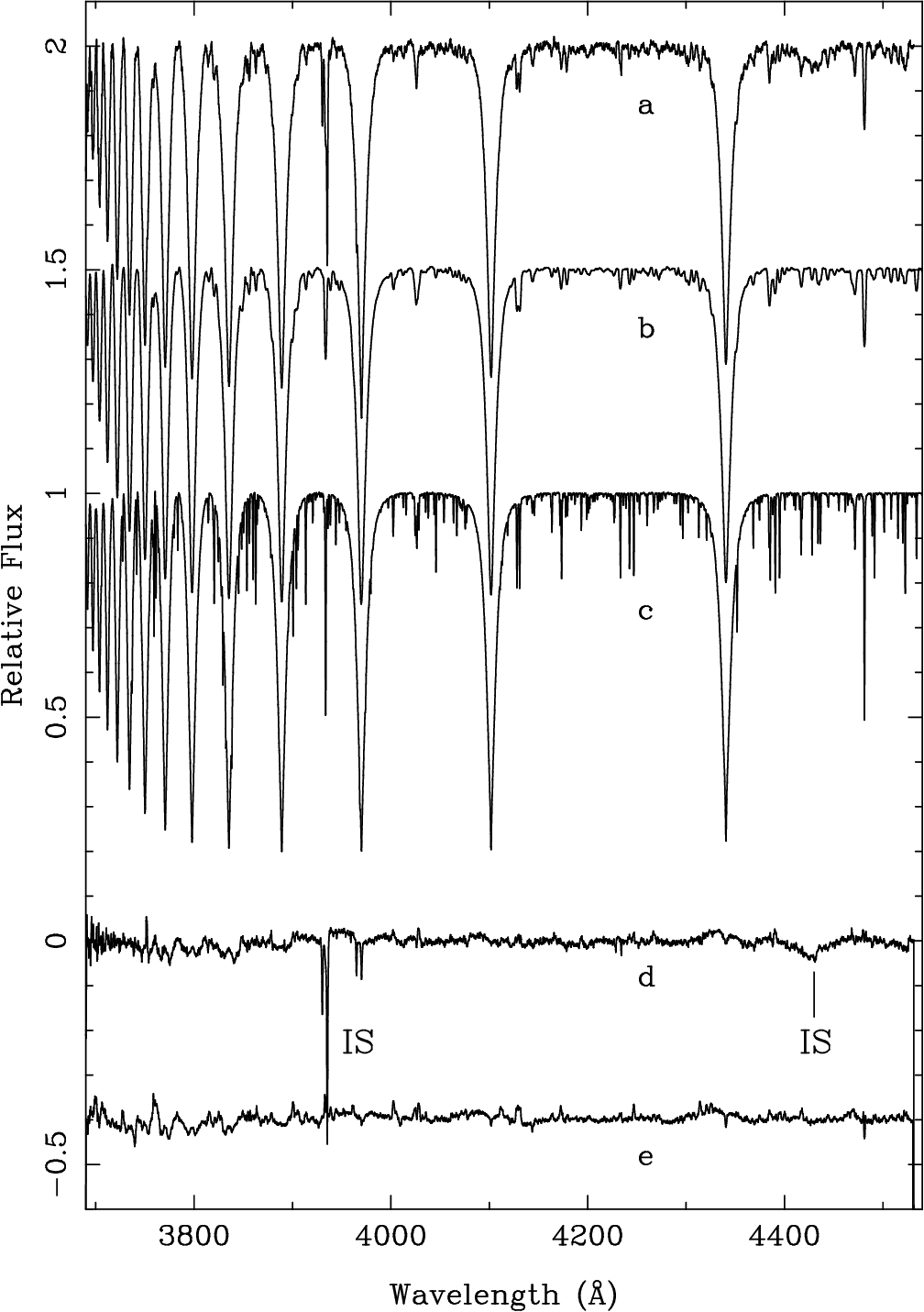}
\figcaption[f10.ps]{(a): The restframe WHT spectrum of V4641 Sgr,
normalized to its continuum and offset upwards by 2.0 units.  (b): A
synthetic spectrum with $T_{\rm eff}=10500$~K, $\log g=3.5$, and
$V_{\rm rot}\sin i=123$ km s$^{-1}$ offset by 1.5.  (c): A synthetic
spectrum with $T_{\rm eff}=10500$~K, $\log g=3.5$, and $V_{\rm
rot}\sin i=1$ km s$^{-1}$, not offset.  (d): The observed spectrum
minus the $T_{\rm eff}=10500$~K, $\log g=3.5$, and $V_{\rm rot}\sin
i=123$ km s$^{-1}$ model, no offset.  
(e).  The observed $\gamma$ Lyr spectrum minus the best fitting model
($T_{\rm eff}=10000$~K, $\log g=3.0$, and
$V_{\rm rot}\sin i=88$ km s$^{-1}$), offset downwards by 0.4 units.
\label{fig4}}
\end{figure*}

We started first with the determination of the rotational velocity.
We computed model spectra for various values of $V_{\rm rot} \sin i$
(using $T_{\rm eff}=10500$~K and $\log g=3.5$) and computed the
minimum reduced $\chi^2$ of the polynomial fit to the difference
spectra, masking out the few anomalous lines.  It is well known that
there is a systematic error in the measurement of $V_{\rm rot}\sin i$
caused by the non-spherical shape of the Roche lobe and by the
variations in the temperature and limb darkening over the Roche
surface.  Therefore we made nightly restframe spectra using the two
WHT spectra from August 12 and the five WHT spectra from August 13,
thus insuring that the restframe spectra are composed of individual
spectra with similarly distorted line profiles.  Fig.\ \ref{rmsfig}a
shows the reduced $\chi^2$ as a function of the input $V_{\rm rot}\sin
i$ for the two restframe spectra.  The minimum reduced $\chi^2$ for
the August 13 spectrum (which has the larger SNR) is for $V_{\rm
rot}\sin i=118$ km s$^{-1}$.  However, the curve is somewhat noisy,
and there is another dip near $V_{\rm rot}\sin i=113$ km s$^{-1}$.
The corresponding curve for the August 12 restframe spectrum has its
minimum at $V_{\rm rot}\sin i=124$ km s$^{-1}$, and another broad
minimum near $V_{\rm rot}\sin i=130$ km s$^{-1}$.

It is somewhat difficult to choose the correct value of $V_{\rm
rot}\sin i$ on the basis of these results.  We repeated the exercise
using phase-specific broadening kernels, computed using the ELC code
\citep{oro00}.  The ELC code uses specific intensities from model
atmosphere computations rather than from blackbodies and a one- or
two-parameter limb darkening law. Thus the distortions in the
broadening kernel due to tidal distortions and to temperature and limb
darkening variations are accounted for in detail.  For the present
problem we used the specific intensities from the Kurucz grid.  We
computed a broadening kernel for each of the two nights by averaging
kernels made for each individual spectrum which went into the
restframe average.  We used a custom made IRAF routine to compute the
convolution.  Fig.\ \ref{rmsfig}b shows the resulting reduced $\chi^2$
vs.\ $V_{\rm rot}\sin i$.  The two curves are much ``cleaner'', and
both have well-defined minima near $V_{\rm rot}\sin i=123$ km
s$^{-1}$.  Fig.\ \ref{fig5} shows the numerical kernels used.  In the
case of the August 13 spectrum, the analytic kernel scaled by 118 km
s$^{-1}$ has almost the same full width at half maximum as the
slightly asymmetric numerical kernel, scaled by 123 km s$^{-1}$.
Thus, to first order, we would expect a correction from 118 to 123 km
s$^{-1}$, which is what we find.  Similarly, the analytic kernel
scaled by 130 km s$^{-1}$ approximately matches the full width at half
maximum of the numerical kernel for August 12, scaled by 123 km
s$^{-1}$.  Thus we adopt $V_{\rm rot}\sin i=123$ km s$^{-1}$.  The
curves in Fig.\ \ref{rmsfig}b start to become steep at $\approx 119$
and 127 km s$^{-1}$, so we adopt a $1\sigma$ error of 4 km s$^{-1}$.

\subsection{Abundance Anomalies}

After determining the rotational velocity we then examined in more
detail the metal lines that were unexpectedly strong.  SYNPLOT can
alter the abundance of any given element and compute an approximate
line profile.  The resulting line profile is reasonably accurate
provided the abundance of the element in question is not altered by
more than a factor of $\approx 3$ from its abundance in the input
model atmosphere (I. Hubeny, private communication).  Since we
currently cannot compute a full grid of 
the fully line-blanketed Kurucz models with
arbitrary abundance patterns, our results should be taken as only
indicative.  We computed models with altered abundances and compared
them to the observed V4641 Sgr spectrum.  In searching for elements
with possible abundance anomalies, we were guided by the work of
\citet{isr99} who found that the abundances of the $\alpha$-process
elements nitrogen, oxygen, magnesium, silicon, and titanium in the
F-star secondary of GRO J1655-40 are 6-10 times solar.  We
concentrated on these same elements in V4641 Sgr.  As a check, we also
computed models for $\gamma$ Lyr with altered abundances.  The results
are given in Table \ref{tabab} and in Figs.\ \ref{plotabund},
\ref{plotabundlyr}, and \ref{abundlow}.  Details for each element are
provided below.

Taken at face value, the LTE Kurucz models indicate that the magnesium
in V4641 Sgr is enhanced by about a factor of five to
seven with respect to solar.
However,
we cannot fit the magnesium lines in $\gamma$ Lyr using the LTE Kurucz
models
(Fig.\ref{plotabundlyr}), which is an indication that a NLTE treatment
is needed.  In Figure \ref{nltemg} we show the Kurucz models and the
Hauschildt models for V4641 Sgr and $\gamma$ Lyr near the Mg II
$\lambda$4481.22 line.  In the case of V4641 Sgr ($T_{\rm eff}=10500$,
$\log g=3.5$), the NLTE treatment makes almost no difference to the strength
of the $\lambda$4481.22 line.  In contrast, the NLTE treatment
makes the $\lambda$4481.22 line much stronger in $\gamma$ Lyr
($T_{\rm eff}=10000$, $\log g=3$).  We conclude the unusual strength of
the $\lambda$4481.22 line in V4641 Sgr is due to an enhancement
of magnesium of a factor of five to seven times solar.

The pair of lines at $\lambda$4128.04 and 4130.58 due to Si II
indicates roughly solar abundance in V4641 Sgr; but as with magnesium,
these Si II lines are not well modelled in $\gamma$ Lyr.  NLTE effects
probably are important for Si II as well.

Perhaps the most convincing case is for oxygen.  The weak O I line at
$\lambda$4368.2 observed in V4641 Sgr requires an oxygen enhancement
of 3 times solar, whereas the solar abundance model for $\gamma$ Lyr
fits the line profile in that star.  For comparison, \citet{bal86}
measured an oxygen abundance in $\gamma$ Lyr of about 1.3 times solar.
The unresolved O I triplet near $\lambda$6454.4 appears in the VLT
spectrum (resolution 3.2~\AA\ FWHM, Fig.\ \ref{abundlow}).  Its
strength suggests the abundance is even a bit more than three times
solar (we did not attempt any further iteration, given the low
resolution of the VLT spectrum).

The stellar Ca II line at $\lambda$3933.66 in V4641 Sgr was resolved
from the interstellar line in the August 12 WHT spectrum.  A calcium
enhancement of twice solar is needed to fit the line.  The Ca II line
in $\gamma$ Lyr is fit by the solar abundance model.

There are no strong titanium lines that are sensitive to the abundance
in the WHT spectrum.  The VLT spectrum (resolution 4~\AA\ FWHM)
contains two Ti II lines that appeared to be much stronger than the
corresponding model lines (Fig.\ \ref{abundlow}).  A titanium
enhancement of a factor of about 10 can explain the profiles of the
$\lambda5226.5$ and $\lambda$5129.2 lines, but not the blend at
$\approx \lambda$5187.
 
There are no strong nitrogen lines in the WHT spectrum.  There is a
line due to N I in the VLT spectrum (resolution 3.2~\AA\ FWHM) which
is anomolous.  A model with a nitrogen enhancement of a factor of 10
solar gives a reasonably good fit to the profile.  We also searched
the WHT and VLT spectra for suitable sulphur lines, but we did not
find any.

To summarize the results of our {\em exploratory} abundance analysis,
we find the following enhancements: nitrogen is $\approx 10$ times
solar (low resolution spectrum only); oxygen is $\approx 3$ times
solar (low and high resolution spectra); magnesium is enhanced by 
about five to seven times solar;
silicon is roughly solar, but
the LTE models may not be adequate; calcium is $\approx 2$ times
solar (high resolution spectrum); and titanium is $\approx 10$ times
solar (low resolution spectrum only).  As we stated above, these
results should be treated with extreme caution.  We do not claim that
the lines we examined are the most suitable ones for an abundance
analysis, nor do we claim the solar metallicity Kurucz models are the
most appropriate.  Clearly we need better data (e.g.\ higher
resolution spectra with lines from more than one ionization stage of
the elements in question) and better models (e.g.\ NLTE
treatment for all of the elements, etc.).

Finally, we repeated the procedure to find the rotational velocity,
using a model with altered oxygen, magnesium, and calcium abundances.
We found no change in the value of $V_{\rm rot}\sin i$.  Fig.\
\ref{fig4} shows the normalized WHT spectrum, the best (normalized)
model, and the difference between the two.  The agreement between the
model and the data is quite good.  The accretion disk is faint; it
contributes less than 1\% of the light.  For comparison, we also show
the difference between the model for $\gamma$ Lyr and the observed
spectrum (spectrum e).

\section{Astrophysical Parameters}\label{impsec}

Now that we have presented the observational facts concerning V4641
Sgr, we proceed to discuss the astrophysical implications of these
observations.

\subsection{Inclination Constraints}

A straightforward upper limit to the inclination comes from the lack
of X-ray eclipses.  As we noted earlier, the source was active as a
relatively weak X-ray source between 1999 February and September and
was detected on numerous occasions by the PCA instrument on {\rm
RXTE}.  C.  Markwardt (private communication) informs us that the
source was detected in X-rays several times near the inferior
conjunction of the secondary star (i.e.\ a {\em spectroscopic} phase
range of $0.75\pm 0.04$).  Assuming the X-rays come from a relatively
small region $R\lesssim 0.01\,R_{\odot}$ centered on the compact
object, one can show $i\le 70.7^{\circ}$ for $Q=1.5$.

In principle, one can determine the inclination by modelling the
so-called ``ellipsoidal'' variations in the light curve of the B-star
companion.  The B-star fills its critical Roche lobe (there is ongoing
mass transfer) and as such is quite distorted.  As the distorted star
moves about in its 2.8173 day orbit, the projected area on the sky
changes, which in turn, gives rise to changes in the observed flux.
The net result is that one observes a light curve with two maxima per
cycle (corresponding to the quadrature phases when the star is seen
``side-on'') and two minima per cycle (when the star is seen
``end-on'').  The amplitude of the light curve and the details of its
shape is a function of the inclination.

In practice, however, there are difficulties which can introduce
systematic errors into the ellipsoidal modelling.  For example, the
accretion disk can contribute a significant amount of flux, thereby
``diluting'' the ellipsoidal light curve observed from the secondary.
From an empirical point of view, the black hole binaries with cool
companions (G, K, or M-type companions) are subject to the biggest
systematic uncertainties since their light curves change from one
observing run to the next \citep{mcc86,oro96,has93,web00}.  On the
other hand, the two black hole binaries with hot companions (GRO
J1655-40 with an F6III companion and 4U 1543-47 with an A2V companion)
have light curves which are quite stable, and it is apparent that the
underlying ellipsoidal light curve of the secondary star dominates in
these two cases.  The light curves of GRO J1655-40  have
been recently modelled and reasonably tight inclination constraints
(to within about 5 degrees at 90\% confidence) have been obtained
(Greene, Bailyn, \& Orosz 2000).  Since V4641 Sgr
contains a hot companion, it seems reasonable to expect that its light
curve will be dominated by the ellipsoidal modulations from the
secondary.

The only well-sampled light curve we have access to at the moment is
the photometric light curve assembled by \citet{gor90}.
Unfortunately, the individual measurements are relatively imprecise
(the mean estimated error is 0.1 mag), and the folded and binned light
curve (Fig.\ \ref{fig3}) has relatively large error bars.
Nevertheless, the ellipsoidal modulation is apparent in the binned
light curve.  For the moment we will limit ourselves to some
representative models and put off the more detailed modelling until
precise CCD light curves in several filter bandpasses become
available.  We used the ELC code \citep{oro00} to compute the models
using specific intensities from the Kurucz grid.  The dashed line in
Fig.\ \ref{fig3} is a Johnson $B$-band model for a lobe-filling star
($T_{\rm mean}=10500$~K) with an inclination of $i=70^{\circ}$ and a
mass ratio of $Q=1.5$.  The amplitude of the model is too small
($\chi^2=35.196$ for the 20 data points).  The solid line in Fig.\
\ref{fig3} is the same model, but with a faint accretion disk added
(the outer radius is 98\% of the Roche lobe radius).  The depth of the
minimum at phase 0.5 is increased by the partial eclipse of the B-star
by the disk.  In this case, the amplitude of the model is much closer
to what is observed, and the model fit is improved ($\chi^2=25.39$ for
the 20 data points).  The model disk contributes about 0.25\% of the
light in the $B$-band, and this would be compatible with the observed
limits (Sec.\ \ref{secsec}).  To find an approximate lower limit to
the inclination, we adjusted the inclination of the model until the
$\chi^2$ of the fit increased by 9.  The mass ratio and the radius of
the disk were not adjusted.  We find $i\gtrsim 60^{\circ}$.

We adopt an inclination range that is uniform in the interval
$60^{\circ}\le i\le 70.7^{\circ}$.  A partial eclipse of the secondary
star is needed to reproduce the relative depths of the minima.  These
conclusions (apart from the upper limit on $i$ which is a function of
$Q$ only) should be treated with caution, given that the light curve
is relatively uncertain.

\subsection{Binary Parameters}

\begin{figure*}[t]
\epsscale{1.0}
\plotone{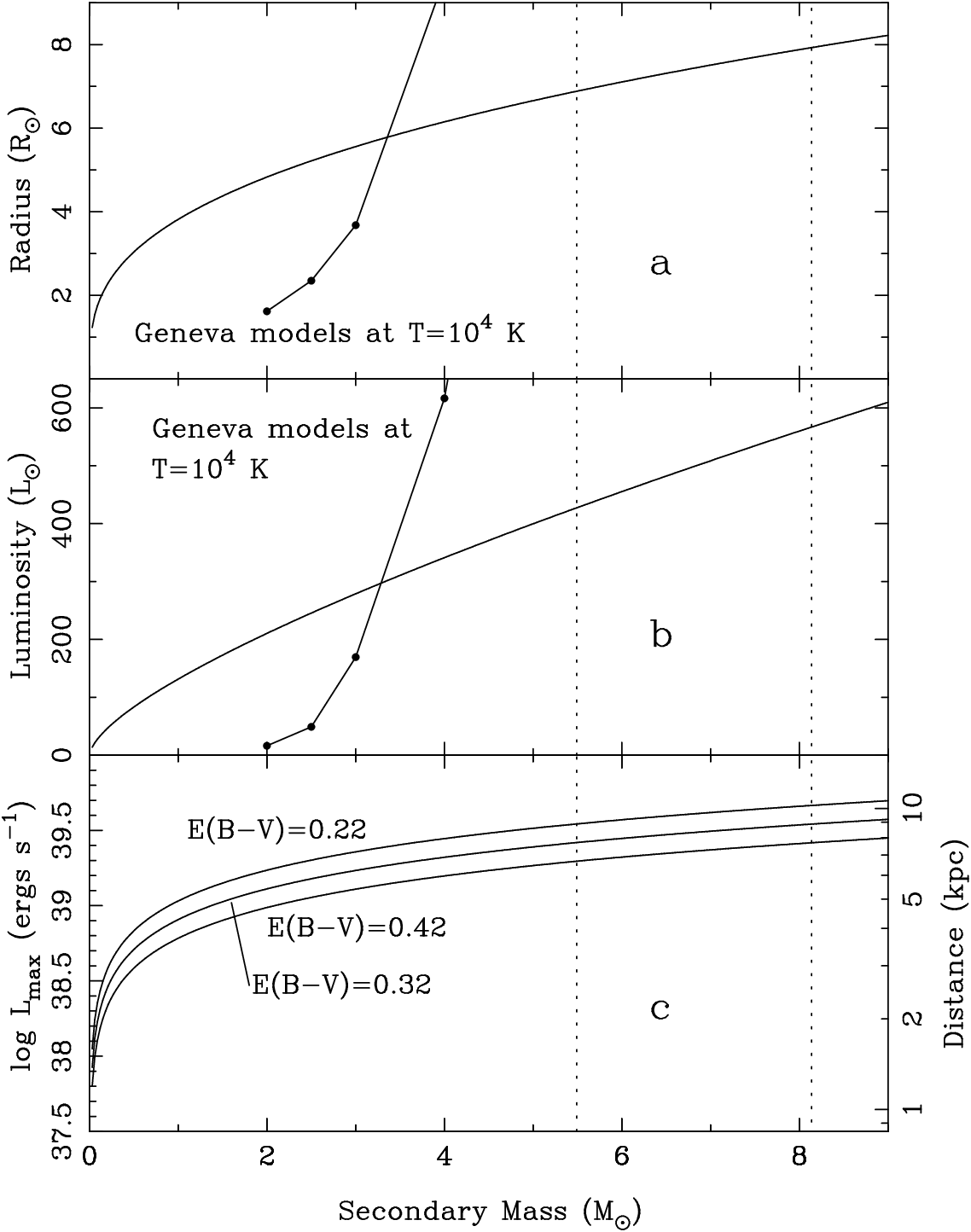}
\figcaption[f11.ps]{Top:  The radius of the secondary star
as a function of its assumed mass.  The points are from the single
star Geneva
stellar evolution models.
Middle:  Similar to the top, except the luminosity of the secondary star
is displayed.
Bottom:  The distance of the source (right $y$-axis scale)
and the peak isotropic X-ray luminosity (left $y$-axis scale)
as a function of the assumed secondary star mass for three
different values of the color excess.  The photometry of the source
in quiescence suggests $E(B-V)=0.32$.
\label{fig6}}
\end{figure*}

For a Roche lobe filling star it is straightforward to show that
\begin{equation}
{V_{\rm rot}\sin i\over K_2}={R_{L_1}(Q)\over a}\left({1+Q\over
Q}\right),
\label{equ1}
\end{equation}
where $Q$ is the mass ratio, $R_{L_1}(Q)$ is the sphere-equivalent
radius of the Roche lobe, and $V_{\rm rot}=2\pi R_{L_1}/P$.  We used
the ELC code to numerically compute $R_{L_1}(Q)$ for $V_{\rm rot}
=123\pm 5$ km s$^{-1}$ and $K_2= 211.0\pm 3.1$ km s$^{-1}$; we find
$Q=1.50\pm 0.13$ (90\% confidence).

We used a simple Monte Carlo code to compute other interesting binary
parameters (e.g.\ the component masses, secondary star radius, etc.)
and their uncertainties using four input quantities, namely the period
($P=2.81730\pm 0.00001$ days, $1\sigma$), the velocity semiamplitude
of the secondary star ($K_2=211.0\pm 3.1$ km s$^{-1}$, $1\sigma$), the
projected rotational velocity of the secondary star ($V_{\rm rot}\sin
i=123\pm 4$ km s$^{-1}$, $1\sigma$), and the inclination
($60.0^{\circ}\le i\le 70.7^{\circ}$, uniform distribution).  The
nominal masses of the black hole and secondary star are
$9.6\,M_{\odot}$ and $6.5\,M_{\odot}$, respectively; the precise
masses and other key results are summarized in Table \ref{tab3}.

\section{Discussion}\label{dissec}

The mass of the compact object in V4641 Sgr ($9.6\,M_{\odot}$) is well
above the maximum mass of a stable neutron star, which is usually
taken to be $\approx 3\,M_{\odot}$ \citep{chi76}.  We therefore
conclude that V4641 Sgr contains a black hole.  V4641 Sgr is also
interesting in several other respects.  The B-star secondary is by far
the most massive, the hottest, and the most luminous secondary of the
ten dynamically confirmed transient black hole binaries.  The total
mass of the system, $16.2\,M_{\odot}$, is also largest among the
dynamically confirmed black hole binaries.  The next most massive
system is V404 Cyg whose total mass is about $12.7\,M_{\odot}$
\citep{sha94}.  The other eight confirmed transient black hole
binaries have total masses which are probably below $10\,M_{\odot}$,
although the uncertainties are large in many cases \citep[and cited
references]{bai98}.

The secondary star in V4641 Sgr appears to be in an unusual
evolutionary state.  To help illustrate this, we plot in Fig.\
\ref{fig6}ab the radius of the secondary and the luminosity of the
secondary as a function of its assumed mass.  Since the density of a
Roche lobe filling star depends only on the orbital period of the
binary to a good approximation, these curves are essentially
independent of the inclination or the mass ratio of the binary.  We
also show the radius and luminosity of {\em single} stars, taken from
the ``Geneva'' stellar evolution models \citep{sch92}.  For our
adopted mass range of $5.49 \le M_2\le 8.14\,M_{\odot}$ (90\%
confidence), the secondary star is both smaller in size and
significantly underluminous compared to single stars of similar
masses.  An underluminous secondary star usually indicates that the
star went through a phase of thermally unstable mass transfer.  A more
detailed discussion of the evolutionary history of this star is beyond
the scope of the current paper.

\begin{figure*}[t]
\epsscale{1.00}
\plotone{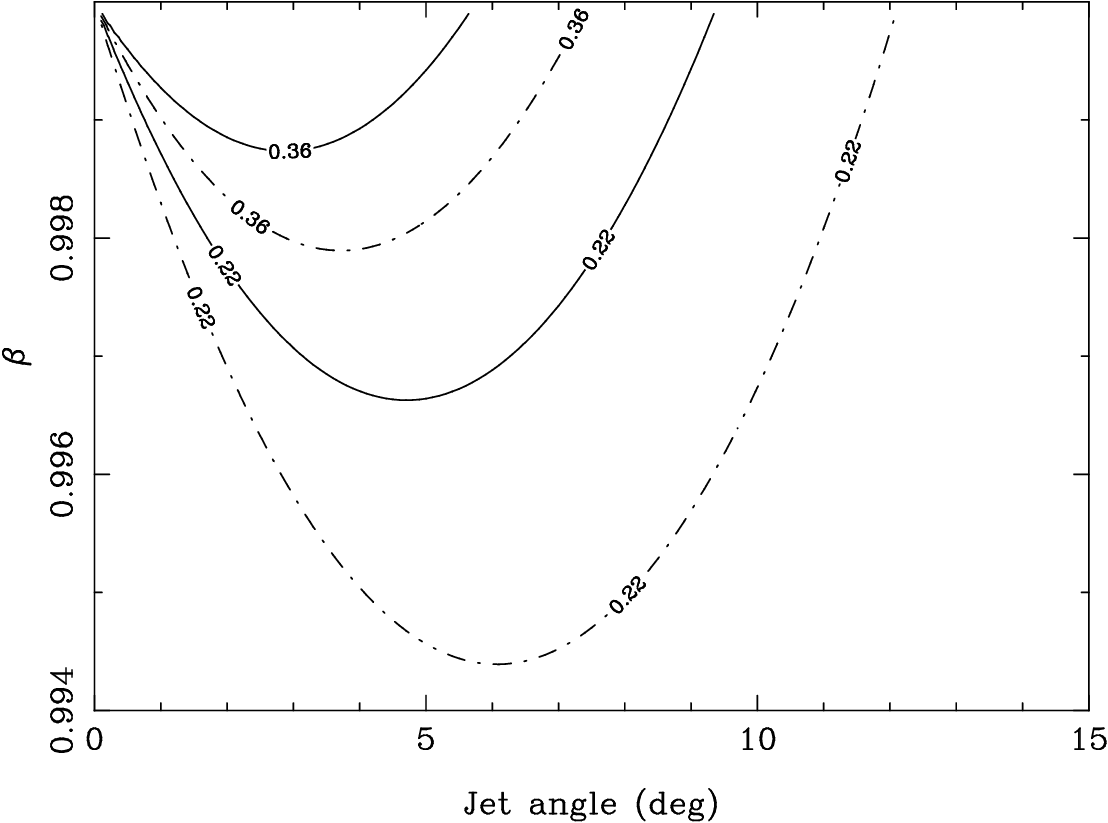}
\figcaption[f12.ps]{A contour plot of the apparent proper motion
in arcseconds per day
of the approaching jet in V4641 Sgr as a function of the jet angle and
the intrinsic jet velocity in units of $c$.  The dash countours are for
a distance of
7.4 kpc and the solid contours are for a distance of 9.6 kpc.
\label{jetcont}}
\end{figure*}

Two remarkable features of V4641 Sgr, which are discussed below, are a
consequence of its great distance.  For an assumed secondary star
mass, we know precisely the corresponding radius.  The corresponding
luminosity is also precisely known since the temperature is well known
from the spectrum.  The distance then follows from the apparent $V$
magnitude and the extinction.  Photometry of the source was obtained
from CTIO 2000 July 23.1 (UT).  We find $B = 14.00 \pm 0.02$ and $V =
13.68 \pm 0.02$.  The intrinsic colors of this star are all close to
zero, hence $E(B-V)=0.32$.  The star is of course variable, so for now
we adopt an uncertainty on the $B$ and $V$ magnitudes of 0.1 mag.
Fig.\ \ref{fig6}c shows the distance as a function of the assumed
secondary star mass.  Assuming $T_{\rm eff}=10500\pm 200$~K and $5.49
\le M_2\le 8.14\,M_{\odot}$ (90\% confidence), we adopt $7.40 \le d
\le 12.31$ kpc (90\% confidence).

As we noted earlier, V4641 Sgr was marginally resolved in the first
radio map made after the 12 crab X-ray flare.  The proper motion of
the jet is uncertain since it is not known when the jet was ejected.
\citet{hje00} give three possibilities for the jet proper motion,
based on known X-ray/radio correlations in other sources:
\begin{itemize}
\item $\mu_{\rm jet}=0.22$ arcseconds per day for an ejection at the
start of the 4.5 crab X-ray flare, which implies $V_{\rm app}\ge
9.47c$ for $d>7.4$ kpc;
\item $\mu_{\rm jet}=0.36$ arcseconds per day for an ejection at the
start of the 12 crab X-ray flare (this is their preferred value),
which implies $V_{\rm app}\ge 15.4c$ for $d>7.4$ kpc;
\item $\mu_{\rm jet}=1.1$ arcseconds per day for an ejection at the
end of the 12 crab flare, which implies $V_{\rm app}\ge 47.7c$ for
$d>7.4$ kpc.
\end{itemize}
For comparison, all of the other Galactic sources with relativistic
jets have apparent velocities of less than $2c$ \citep{mir99}.  
Indeed, this source might be characterized as a galactic
``microblazar'', since the jet velocity is similar to what is typcially
seen in blazars \citep{jor01}.  It is worthwhile to point out that
there are other interpretations of the radio observations.  For example, 
\citet{cha01} note that the long-lived radio remnant seen in
Fig.\ 5c of \cite{hje00} did not show any movement between 1999 September 16
and 24.  This raises the possibility that the radio emission was due
to an interaction of ejecta with the surrounding interstellar medium at a
distance of $\approx 0.25d$ AU from the source, where the distance $d$ is
measured in parsecs. 
At a distance of 9.6 kpc, an angular size of 0.25 arcseconds corresponds
to a physical size of 2400 AU.  The corresponding light crossing time
is 13.9 days, so the jet would have had to been ejected on $\approx$
September 2 at the latest in order to have interacted with matter
at a distance of 2400 AU.  Optical observers coordinated through
the VSNet noted possible activity as
early as August 
8\footnote{http://www.kusastro.kyoto-u.ac.jp/vsnet/Mail/vsnet/msg01879.html},
so a jet ejection as early as September 2 cannot be ruled out.

The
apparent proper motion of the approaching jet depends on the intrinsic
jet velocity $\beta=v/c$, jet angle $\theta$, and the distance $d$
\citep{mir99}
\begin{equation}
\mu_a={\beta\sin\theta\over 1-\beta\cos\theta}{c\over d}.
\end{equation}
Fig.\ \ref{jetcont} shows a contour plot of $\mu_a$ as a function of
$\theta$ and $\beta$ for assumed distances of $d=7.4$ kpc (our lower
limit at 90\% confidence) and $d=9.6$ kpc (our preferred value).  A
proper motion of $\mu_a>0.22$ arcseconds per day requires $\beta>
0.9945$ and $\theta < 12^{\circ}$ for $d=7.4$ kpc.  The bulk Lorentz
factor in this case is $\Gamma=(1-\beta^2)^{-1/2}>9.5$.  The
combination of $d=9.6$ kpc and $\mu_a=0.36$ arcseconds per day gives
$\beta> 0.999$ ($\Gamma>22.4$) and $\theta < 6^{\circ}$.  In any case,
the bulk motions were extremely relativistic, and the jet in V4641 Sgr
was nearly aligned with the line of sight.  On the other hand, the
large inclination indicates that the orbital plane is nearly
perpendicular to the line of sight.  If the jet is aligned with the
spin vector of the black hole, then it seems likely that the spin
vector of the black hole is misaligned (by a wide margin) with the
orbital angular momentum vector.  The inclination of the transient jet
observed during the 1994 outburst of GRO J1655-40 was about
$85^{\circ}$ \citep{hje95}, whereas the orbital inclination is close
to $70^{\circ}$ \citep{oro97,vdh98,gre00}, indicating a misalignment
of about $15^{\circ}$.  Thus if the inclination of the jet can be
taken as the inclination of the black hole spin vector, then
misaligned black hole spin vectors may be a common feature of these
systems.

The peak {\em isotropic} X-ray luminosity observed during the large
X-ray flare in the 2-10 keV band is $L_{\rm max}=5\times
10^{36}(d/0.4~{\rm kpc})^2$ erg s$^{-1}$ \citep{smi99}.  Using our
adopted distance we find $\log L_{\rm max}=39.46\pm 0.23$ (90\%
confidence) for the 2-10 keV band.  The luminosity in the 10-100 keV
band is comparable to the luminosity in the 2-10 keV band, so the bolometric
luminosity will be a factor of $\approx 2$ higher.  Thus, rather than the
main outburst being substantially sub-Eddington, as initially thought
\citep{int00}, the outburst was actually super-Eddington, since $\log
L_{\rm edd}= 39.10$ for a mass of $9.61\,M_{\odot}$.  The bright soft
X-ray transient 4U 1543-47 had a peak X-ray flux (1.5-3.8 keV) that
was at least three times the Eddington limit \citep{oro98}, so the
super-Eddington peak X-ray flux observed for V4641 Sgr is by no means
unique.

\citet{fen00} have recently shown that the peak X-ray flux $P_X$
observed in the outbursts of the black hole transients is positively
correlated with the peak observed radio flux $P_R$.  If all black hole
X-ray transients produce a mildly relativistic jet (five sources have
been resolved at radio wavelengths), then the existence of the
$P_X$/$P_R$ correlation implies that on average the bulk Lorenz
factors in the jets are relatively small (e.g.\ $\Gamma\lesssim 5$)
since very large values of $\Gamma$ would result in many systems with
radio emission beamed out of the line of sight and hence unobservable
\citep{fen00}.  These authors point out that there is nothing unusual
about V4641 Sgr in terms of its peak X-ray and peak radio flux.  The
radio emission observed in V4641 Sgr was strongly beamed, and the fact
that V4641 Sgr sits on the $P_X/P_R$ correlation suggests that the
X-rays may also have been beamed and that the peak outburst luminosity
was not necessarily super-Eddington.  X-ray beaming may also be
responsible for the super-Eddington outburst of 4U 1543-47 and for the
apparently large numbers (twelve or more) of the binary sources in M82
and Cen A with luminosities of up to $10^{41}$ ergs s$^{-1}$
\citep{pre00}.  These sources may be $\approx 10\,M_{\odot}$ black
holes with beamed X-ray emission rather than $\approx 100\,M_{\odot}$
black holes emitting isotropically at the Eddington limit.

Finally, we note that we can make an independent estimate of the distance 
if we assume that
the large observed radial velocity of the binary (Table
\ref{tab2}) is due entirely to
differential galactic rotation.  
For a source at a galactic longitude of
$\ell=6.77$ and galactic latitude of $b=-4.79$ we find $d>7$ kpc using the
rotation curve given in \citet{fic89} and  the 
standard IAU rotation constants of $R_0=8.5$ kpc and
$\Theta_0=220$ km s$^{-1}$.  This lower limit is in good agreement with
the lower limit we derived above.
The space motion of the binary relative  to
its local standard of rest $V_{\rm corr}$  can be computed from the distance
we derived above
and the observed $\gamma$ velocity.  However, $V_{\rm corr}$ 
is unfortunately rather uncertain owing
to the binary's close proximity to the galactic center.  The distribution of
$V_{\rm corr}$ values computed using the simple Monte Carlo code mentioned
above is 
roughly uniform in the range $-80\le V_{\rm corr}\le 80$ km s$^{-1}$.

Although our basic picture of V4641 Sgr is secure, there are many
useful follow-up observations that can be done.  The photometric light
curve has relatively large errors; and light curves with greater
statistical precision and for several bandpasses should be obtained.
Our tentative conclusion is that the secondary star is partially
eclipsed by the disk.  It should be easy to confirm this with CCD
light curves.  The derived properties of the binary (e.g.\ component
masses, etc.)  are somewhat sensitive to the value of $V_{\rm rot}\sin
i$.  Thus another measurement of the rotational velocity of the B-star
secondary should be obtained, preferably near the time of the inferior
conjunction of the B-star, when the systematic bias introduced by the
tidal distortion of the secondary star is minimized.  Finally, since
V4641 Sgr is so bright, one can easily obtain a high resolution, high
signal-to-noise spectrum and do a more detailed abundance
analysis. This was done for the black hole binary GRO J1655-40 by
\citet{isr99} who reported a large overabundance of oxygen, magnesium,
silicon, and sulphur, which they attributed to the capture of
supernova ejecta from the progenitor of the present-day black hole.
We find that the $\alpha$-process elements nitrogen, oxygen, calcium,
and titanium appear to be overabundant with respect to solar in V4641
Sgr.  However, we do not find the same levels of enrichment in V4641
Sgr that were found for GRO J1655-40.  The amounts of the various
elements synthesized in a supernova explosion depend on the mass of
the progenitor star and on the energy of the explosion \citep{nom00}.
Thus the differences in the levels of enrichment of the
$\alpha$-process elements in GRO J1655-40 and V4641 Sgr may reflect
the differences in the supernovae progenitors.

\section{Summary}\label{sumsec}

We have presented the results of our spectroscopic campaign on the
fast X-ray transient and superluminal jet source V4641 Sgr.  The
spectroscopic period is $P_{\rm spect}=2.81678\pm 0.00056$ days and
the radial velocity semiamplitude is $K_2=211.0\pm 3.1$ km s$^{-1}$.
The optical mass function is $f(M)=2.74\pm 0.12\,M_{\odot}$.  For the
secondary star we measure $T_{\rm eff}=10500 \pm 200$~K, $\log
g=3.5\pm 0.1$, and $V_{\rm rot}\sin i=123\pm 4$ km s$^{-1}$ (1$\sigma$
errors).

The photometric period measured from an archival photographic light
curve is $P_{\rm photo}=2.81730 \pm 0.00001$ days.  The light curve
folded on the photometric phase resembles an ellipsoidal light curve.
Modelling of this light curve indicates a high inclination angle of
$i\gtrsim 60^{\circ}$.  The lack of X-ray eclipses implies an upper
limit to the inclination of $i\le 70.7^{\circ}$.  Using these
inclination limits and the above spectroscopic parameters, we find a
compact object mass in the range $8.73\le M_1 \le 11.70\,M_{\odot}$
(90\% confidence).  This mass range is well above the maximum mass of
a stable neutron star and we conclude that V4641 Sgr contains a black
hole.

The secondary is a late B-type star which has evolved off the main
sequence.  It is in a peculiar evolutionary state since both its
radius and its luminosity are much smaller than the corresponding
values for single stars with a similar mass.

Finally, we find a distance in the range $7.40\le d \le 12.31$ kpc
(90\% confidence), which is at least a factor of $\approx 15$ larger
than the initially assumed distance of $\approx 500$ pc.  The peak
X-ray luminosity was super-Eddington, and the apparent expansion
velocity of the radio jet was $\gtrsim 9.5c$.

\acknowledgments

We thank the unknown referee for several helpful comments which improved
this paper, and Peter Hauschildt for providing the model spectra.
We acknowledge the use of the PERIOD package developed by Vik Dhillon
from the Starlink Software Collection, and the SYNPLOT spectrum
synthesis programs written by Ivan Hubeny.  It is a great pleasure to
thank the numerous people at the four different observatories who made
this project a success.  P. Berlind and M. Calkins were the FLWO
observers.  M. Chadid and D.  Hutsemekers were the support astronomers
for the VLT observations.  P. Leisy was the support astronomer for the
NTT.  Leisy and the rest of the NTT team headed by O. Hainaut went
beyond the call of normal duty by reconfiguring EMMI on short notice.
N. Walton was the support astronomer at the WHT.  We acknowledge
useful discussions with Ivan Hubeny, Vitaly Goranskij, Craig
Markwardt, Peter Hauschildt, Frank Verbunt, Norbert Langer, Marten van
Kerkwijk, and the late Bob Hjellming, who will be greatly missed.

\clearpage

\begin{deluxetable}{lr}
\tablecaption{Orbital Parameters for V4641 Sgr. \label{tab2}}
\tablewidth{0pt}
\tablehead{
\colhead{Parameter} & \colhead{Value}
}
\startdata
Orbital period, spectroscopic (days) & $2.81678\pm 0.00056$ \cr
Orbital period, photometric (days)   & $2.81730\pm 0.00001$ \cr
$K_2$ velocity (km s$^{-1}$)         & $211.0\pm 3.1$    \cr
$\gamma$ velocity (km s$^{-1}$)       & $107.4 \pm 2.9$     \cr
$T_0$, spectroscopic\tablenotemark{a} ~(HJD 2,451,000+) & $442.523\pm 0.052$ \cr
$T_0$, photometric\tablenotemark{b} ~(HJD 2,447,000+) & $707.4865\pm 0.0038$ \cr
Mass function ($M_{\odot}$)  & $2.74\pm 0.12$ \cr
$V_{\rm rot}\sin i$ (km s$^{-1}$) & $123\pm 4$
\enddata
\tablecomments{All quoted uncertainties are $1\sigma$.}
\tablenotetext{a}{The time of the maximum radial velocity of the 
secondary star.}
\tablenotetext{b}{The time of the deeper photometric minimum,
corresponding to the superior conjunction of the secondary star.}
\end{deluxetable}

\begin{deluxetable}{rccccc}
\tablecaption{Results for abundance analysis. \label{tabab}}
\tablewidth{0pt}
\tablehead{
\colhead{Line} & \colhead{Spectral} &\colhead{V4641 Sgr}    
&  \colhead{solar abundance} & 
\colhead{adopted model} & \colhead{abundance} \\
\colhead{ } & \colhead{resolution (\AA)} & \colhead{$W_{\lambda}$(\AA)} 
 &
\colhead{model $W_{\lambda}$ (\AA)} 
& \colhead{$W_{\lambda}$ (\AA)} & \colhead{($\times$solar)} 
}
\startdata
N I $\lambda6483.7$ & 3.20 & $0.115\pm 0.007$ &    0.011 & 0.078 & 10\cr
                    &      &                  &          &       &   \cr
O I $\lambda4368.2$ & 0.58 & $0.099\pm 0.002$ &    0.033 & 0.056 & 3 \cr
                    &      &                  &          &       &   \cr
O I $\lambda6453.6$ & 3.20 &                  &          &       &   \cr
O I $\lambda6454.4$ & 3.20 & $0.226\pm 0.005$ &    0.097 & 0.115 & 3 \cr
O I $\lambda6456.0$ & 3.20 &                  &          &       &   \cr
                    &      &                  &          &       &   \cr
Mg II $\lambda4481.2$ &0.58& $0.455\pm 0.010$ &    0.310 & 0.476 & 7 \cr
                    &      &                  &          &       &   \cr
Ca II $\lambda3933.7$ &0.58& $0.573\pm 0.010$ &    0.424 & 0.550 & 2 \cr
                    &      &                  &          &       &   \cr
Ti II $\lambda5129.2$ &4.00& $0.085\pm 0.005$ &    0.019 & 0.066 & 10\cr
                    &      &                  &          &       &   \cr
Ti II $\lambda5185.9$ & 4.00 &                  &          &       &   \cr
Ti II $\lambda5188.7$ & 4.00 & $0.083\pm 0.012$ &    0.129 & 0.247 & 1 \cr
                    &      &                  &          &       &   \cr
Ti II $\lambda5226.5$ & 4.00 & $0.110\pm 0.005$ &    0.076 & 0.128 & 10
\enddata
\tablecomments{All quoted uncertainties are $1\sigma$.}
\end{deluxetable}

\begin{deluxetable}{lr}
\tablecaption{Astrophysical Parameters for V4641 Sgr. \label{tab3}}
\tablewidth{0pt}
\tablehead{
\colhead{Parameter} & \colhead{Value}
}
\startdata
Black hole mass ($M_{\odot}$) & $9.61^{+2.08}_{-0.88}$  \cr
Secondary star mass ($M_{\odot}$) & $6.53^{+1.6}_{-1.03}$      \cr
Total mass ($M_{\odot}$) & $16.19^{+3.58}_{-1.94}$      \cr
Mass ratio               &  $1.50\pm 0.13$ \cr
Orbital separation ($R_{\odot}$) & $21.33^{+1.25}_{-1.02}$ \cr
Secondary star radius ($R_{\odot}$) & $7.47^{+0.53}_{-0.47}$  \cr
Secondary star luminosity ($L_{\odot}$)\tablenotemark{a} & $610^{+122}_{-104}$  \cr
Distance (kpc)\tablenotemark{b}           & $9.59^{+2.72}_{-2.19}$    \cr
$\log L_x$ (erg s$^{-1}$)\tablenotemark{c} & $39.46^{+0.23}_{-0.20}$
\enddata
\tablecomments{The quoted errors are all 90\% confidence.}
\tablenotetext{a}{Assuming $T_{\rm eff}=10500\pm 200$~K, $1\sigma$.}
\tablenotetext{b}{Assuming $E(B-V)=0.32\pm 0.10$, $1\sigma$ 
and $A_V=3.1E(B-V)$.}
\tablenotetext{c}{The peak X-ray luminosity (2-10 keV), given
by $5\times 10^{36}(d/4~{\rm kpc})^2$ erg s$^{-1}$, \protect\citep{smi99}.}
\end{deluxetable}

\end{document}